\documentclass{aa}

\usepackage{astron}
\usepackage{psfig}

\begin{document}
%
\def\lfir{$L_{\rm FIR}$}
\def\mabs{M$_{\rm abs}$}
\def\etal{et al.}
\def\cbeta{$c_{\rm H\beta}$}
\def\av{A$_{\rm v}$}
\def\flam{$F_{\lambda}$}
\def\ilam{$I_{\lambda}$}
\def\teff{\ifmmode T_{\rm eff} \else $T_{\mathrm{eff}}$\fi}
\def\lg{$\log g$}
\def\feh{$\mathrm{[Fe/H]}$}
\def\mh{$\mathrm{[M/H]}$}
\def\ltsima{$\buildrel<\over\sim$}
\def\lsim{\lower.5ex\hbox{\ltsima}}

\newcommand{\hii}{H~{\sc ii}}
\newcommand{\ha}{\ifmmode {\rm H}\alpha \else H$\alpha$\fi}
\newcommand{\hb}{\ifmmode {\rm H}\beta \else H$\beta$\fi}
\newcommand{\lya}{Lyman-$\alpha$}
\newcommand{\hei}{He~{\sc i}}
\newcommand{\Hei}{He~{\sc i} $\lambda$4471}
\newcommand{\heii}{He~{\sc ii}}
\newcommand{\Heiiuv}{He~{\sc ii} $\lambda$1640}
\newcommand{\Heiiopt}{He~{\sc ii} $\lambda$4686}
\newcommand{\qh}{\ifmmode q({\rm H}) \else $q({\rm H})$\fi}
\newcommand{\qhe}{\ifmmode q({\rm He^0}) \else $q({\rm He^0})$\fi}
\newcommand{\qhep}{\ifmmode q({\rm He^+}) \else $q({\rm He^+})$\fi}
\newcommand{\Qh}{\ifmmode Q({\rm H}) \else $Q({\rm H})$\fi}
\newcommand{\Qhe}{\ifmmode Q({\rm He^0}) \else $Q({\rm He^0})$\fi}
\newcommand{\Qhep}{\ifmmode Q({\rm He^+}) \else $Q({\rm He^+})$\fi}
\newcommand{\Qhtwo}{\ifmmode Q({\rm LW}) \else $Q({\rm LW})$\fi}
\newcommand{\qrathe}{\ifmmode q({\rm He^0})/q({\rm H}) \else $q({\rm He^0})/q({\rm H})$\fi}
\newcommand{\qrathep}{\ifmmode q({\rm He^+})/q({\rm H}) \else $q({\rm He^+})/q({\rm H})$\fi}
\newcommand{\Qrathe}{\ifmmode Q({\rm He^0})/Q({\rm H}) \else $Q({\rm He^0})/Q({\rm H})$\fi}
\newcommand{\Qrathep}{\ifmmode Q({\rm He^+})/Q({\rm H}) \else $Q({\rm He^+})/Q({\rm H})$\fi}
\newcommand{\Qhave}{\ifmmode \bar{Q}({\rm H}) \else $\bar{Q}({\rm H})$\fi}
\newcommand{\Qheave}{\ifmmode \bar{Q}({\rm He^0}) \else $\bar{Q}({\rm He^0})$\fi}
\newcommand{\Qhepave}{\ifmmode \bar{Q}({\rm He^+}) \else $\bar{Q}({\rm He^+})$\fi}
\newcommand{\Qhtwoave}{\ifmmode \bar{Q}({\rm H}_2) \else $\bar{Q}({\rm H}_2)$\fi}
\newcommand{\Qratheave}{\ifmmode \bar{Q}({\rm He^0})/\bar{Q}({\rm H}) \else $\bar{Q}({\rm He^0})/\bar{Q}({\rm H})$\fi}
\newcommand{\Qrathepave}{\ifmmode \bar{Q}({\rm He^+})/\bar{Q}({\rm H}) \else $\bar{Q}({\rm He^+})/\bar{Q}({\rm H})$\fi}

\def\micron{$\mu$m}
\def\kms{km s$^{-1}$}
\def\kmsmpc{km s$^{-1}$ Mpc$^{-1}$}
\def\cmc{cm$^{-3}$}
\def\erg{ergs s$^{-1}$ cm$^{-2}$ \AA$^{-1}$}
\def\ergs{ergs s$^{-1}$}
\def\ergscm{ergs s$^{-1}$ cm$^{-2}$}
\def\msun{\ifmmode M_{\odot} \else M$_{\odot}$\fi}
\def\zsun{\ifmmode Z_{\odot} \else Z$_{\odot}$\fi}
\def\lsun{\ifmmode L_{\odot} \else L$_{\odot}$\fi}

\def\mup{\ifmmode M_{\rm up} \else M$_{\rm up}$\fi}
\def\mlow{\ifmmode M_{\rm low} \else M$_{\rm low}$\fi}
\def\ubvetc{(UBV)$_J$\-(RI)$_C$\- JHKLL$^\prime$M}
\def\basel {{\it B}a{\it S}e{\it L}}
\def\was{CMT$_1$T$_2$}
%
\def\aap{A\&A}
\def\aaps{A\&AS}
\def\aas{A\&AS}
\def\aj{AJ}
\def\apj{ApJ}
\def\apjl{ApJ}
\def\apjs{ApJS}
\def\mnras{MNRAS}
\def\pasp{PASP}
%


\title{On the properties of massive Population III stars and metal-free stellar
populations
}

\author{Daniel Schaerer \inst{1}}

\offprints{D. Schaerer, schaerer@ast.obs-mip.fr}

 \institute{Observatoire Midi-Pyr\'en\'ees, Laboratoire d'Astrophysique, UMR
 5572, 14, Av.  E. Belin, F-31400 Toulouse, France }

\date{Received 2 july 2001 / Accepted 13 november 2001}

\titlerunning{Population III stars and stellar populations}

\abstract{see below
}

\maketitle

\section*{Abstract}
We present realistic models for massive Population III stars 
and stellar populations based on non-LTE model atmospheres,
recent stellar evolution tracks and up-to-date evolutionary synthesis
models, with the aim to study their spectral properties, including their 
dependence on age, star formation history, and IMF.

A comparison of plane parallel non-LTE model atmospheres
and comoving frame calculations shows that even in the presence of some
putative weak mass loss, the ionising spectra of metal-free 
populations differ little or negligibly from those obtained using plane 
parallel non-LTE models.
As already discussed by Tumlinson \& Shull (2000), the main salient
property of Pop III stars is their increased ionising flux, especially
in the He$^+$ continuum ($>$ 54 eV).

The main result obtained for {\em individual Pop III stars} is the following:
  Due to their redward evolution off the zero age main sequence (ZAMS)
  the spectral hardness measured by the He$^+$/H ionising flux is decreased 
  by a factor $\sim$ 2 when averaged over their lifetime.
  If such stars would suffer strong mass loss, their spectral appearance
  could, however, remain similar to that of their ZAMS position.

The main results regarding {\em integrated stellar populations} are:
\begin{itemize}
\item For young bursts and the case of a constant SFR, nebular continuous
  emission --- neglected in previous studies ---  dominates the spectrum 
  redward of Lyman-$\alpha$ 
   if the escape fraction of ionising photons out of the considered region 
       is small or negligible.
  In consequence predicted emission line equivalent widths are considerably
  smaller than found in earlier studies, whereas the detection of the continuum
  is eased. 
  Nebular line and continuous emission strongly affect the broad band photometric
  properties of Pop III objects.
\item Due to the redward stellar evolution and short lifetimes of the most massive
  stars, the hardness of the ionising spectrum decreases rapidly, leading to the
  disappearance of the characteristic \heii\ recombination lines after $\sim$ 3
  Myr in instantaneous bursts.
\item \Heiiuv, \ha\ (and other) line luminosities usable as indicators of 
  the star formation rate are given for the case of a constant SFR.
  For obvious reasons such indicators depend strongly on the IMF.
\item Due to an increased photon production and reduced metal yields,
   the relative efficiency of ionising photon energy to heavy element rest 
  mass production, $\eta$, 
  of metal-poor and metal-free populations is increased by factors of 
  $\sim$ 4 to 18 with respect to solar metallicity and for ``standard'' IMFs.
\item The lowest values of $\eta \sim$ 1.6 -- 2.2 \% are obtained for IMFs
  exclusively populated with high mass stars (\mlow\ $\ga$ 50 \msun). If correct,
  the yields dominated by pair creation SNae then predict large overabundances
  of O/C and Si/C compared to solar abundance ratios.
\end{itemize}

Detailed results are given in tabular form and as fit formulae for 
easy implementation in other calculations. The predicted spectra 
will be used to study the detectability of Pop III galaxies and to 
derive optimal search strategies for such objects.

\section{Introduction}
\label{s_intro}

Important advances have been made in recent years on the modeling
of the first stars and galaxies forming out of pristine matter 
--- so called Population III (Pop III) objects ---
in the early Universe (see e.g.\ the proceedings of Weiss \etal\ 2000
and Umemura \& Susa 2001).
Among the questions addressed (see also the reviews of Loeb \& Barkana 2001
and Barkana \& Loeb 2001) are 
the star formation process
and the initial mass function (IMF) of the first stars 
(Tegmark \etal\ 1997, Abel \etal\ 1998, 2000, Bromm \etal\ 1999, 
Nakamura \& Umemura 1999, 2001, Omukai \& Palla 2001), 
their effect of energy injection and supernovae (SN)
(MacLow \& Ferrara 1999, Ciardi \etal\ 2000),
their role on cosmic reionisation (Haiman \& Loeb 1997,
Gnedin \& Ostriker 1997, Tumlinson \& Shull 2000, 
Cojazzi \etal\ 2000, Ciardi \etal\ 2001),
metal enrichment of the IGM and other signatures
of early chemical evolution (Ferrara \etal\ 2000, Abia \etal\ 2001),
and dust formation (Todini \& Ferrara 2000).

Extensive sets of recent stellar evolution and nucleosynthesis
calculations for zero metallicity stars covering a wide range of
stellar masses have also become available
recently (Marigo \etal\ 2000, Feij\'oo 1999, Desjacques 2000, 
Woosley \& Weaver 1995, Heger \etal\ 2000, 2001, Umeda \etal\ 2000).
 
Tumlinson \& Shull (2000, hereafter TS00) have recently pointed out 
the exceptionally strong He$^+$ ionising flux of massive ($M \ga$ 40 \msun) 
Pop III stars, which must be a natural consequence of their compactness, 
i.e.\ high effective temperatures, and non-LTE effects in their atmospheres
increasing the flux in the ionising continua.
As a consequence strong \heii\ recombination lines such as \Heiiuv\ or 
\Heiiopt\ are expected; together with AGN a rather unique feature of 
metal-free stellar populations, as discussed by TS00 and Tumlinson 
\etal\ (2001, hereafter TGS01).
Instead of assuming a ``standard'' Population I like Salpeter IMF 
and ``normal'' stellar masses up to 
100 \msun\ as TS00, 
Bromm \etal\ (2001) have considered stars with masses 
larger than 300 \msun, which may form according to some recent hydrodynamical
models (Abel \etal\ 1998, Bromm \etal\ 1999, Nakamura \& Umemura 2001).
An even stronger ionising flux and stronger H and \heii\ emission lines
was found.

Some strong simplifying assumptions are, however, made in the calculations of 
TS00, TGS01, and Bromm \etal\ (2001).
\begin{itemize}
\item[{\em 1)}] All stars are assumed to be on the zero age main sequence, 
   i.e.\ stellar evolution proceeding generally to cooler temperatures is neglected
   \footnote{A simple estimate of this effect is given by TGS01.}.
\item[{\em 2)}] Bromm \etal\ (2001) take the hardness of the stellar spectrum of 
   the hottest (1000 \msun) star as representative for all stars with masses down 
   to 300 \msun. This leads in particular to an overestimate of the \heii\
   recombination line luminosities.
\item[{\em 3)}] None of the studies include nebular continuous emission, which 
   cannot be neglected for metal-poor objects with such strong ionising fluxes.
   This process increases significantly the total continuum flux at wavelengths redward 
   of Lyman-$\alpha$ and leads in turn to reduced emission line equivalent widths.
\item[{\em 4)}] A single, fixed, IMF is considered only by TS00 and TGS01.
   In view of the uncertainties on this quantity it appears useful to explore
   a wider range of IMFs.
\end{itemize}

The models presented here relax all the above assumptions and allow
the exploration of a wide parameter space in terms of stellar tracks 
(including also alternate tracks with strong mass loss), 
IMF, and star formation history (instantaneous bursts or constant 
star formation rate (SFR)).
For this aim we provide in particular the main results in tabulated
form and as fit formulae.

First we examine the influence of several poorly constrained physical 
processes (mass loss, non coherent electron scattering etc.) on the predicted
stellar fluxes of Pop III stars using non-LTE atmosphere models.
Combined with metal-free stellar tracks these predictions are then
introduced to our evolutionary synthesis code (Schaerer \& Vacca 1998)
updated to calculate the nebular properties (H and He emission lines and
nebular continuum) for metal-free gas. 
In this manner we examine the properties of both
individual Pop III stars and integrated stellar populations.

Calculations of the relative photon to metal production of stellar populations
(e.g.\ Madau \& Shull 1996) have proven useful for a variety of
studies ranging from estimates of the stellar contribution to the UV
background radiation (Madau \& Shull 1996), over studies of metals in the
IGM, to SN rates in the early Universe and their detectability 
(Miralda-Escud\'e \& Rees 1997).
To improve on such calculations based on solar metallicity models
we also calculate the heavy element production from metal-poor and metal-free
populations including contributions from type II SN and possible pair 
creation SN from very massive stars (cf.\ Ober \etal\ 1983, Heger \etal\
2000).

It is the hope that our models should in particular help to
examine more precisely the properties of the first luminous objects
in the Universe in preparation of their first direct observation
(cf.\ TGS01, Oh \etal\ 2001a, Schaerer \& Pell\'o 2001,
Pell\'o \& Schaerer 2001).

The paper is structured as follows:
The main model ingredients are summarised in Sect.\ \ref{s_models}.
In Sect.\ \ref{s_single} we discuss the properties of individual Pop III 
stars. The properties of integrated stellar populations, their spectra, 
photon fluxes, metal production etc.\ are presented in Sect.\ \ref{s_pops}.
Our main results are summarised in Sect.\ \ref{s_summary}.

\section{Model ingredients}
\label{s_models}

\subsection{Atmosphere models}
As well known (e.g.\ Mihalas 1978) strong departures from local
thermodynamic equilibrium (LTE) occur in the atmospheres of hot stars.
In addition, for the most massive stars it is of interest to examine
the possible effect of mass loss on their emergent spectra.
Albeit presumably small, mass loss --- due to radiation pressure
or pulsational instabilities (Kudritzki 2000, Baraffe \etal\ 2000)
--- potentially affects the spectral energy distribution
(SED), especially in the ionising continua, as known for Pop I stars (Gabler \etal\
1989, Schaerer \& de Koter 1997).
For stars with negligible ionising fluxes, the remainder of the SED
of interest here (rest-frame UV and optical) is well described by 
plane parallel LTE atmospheres.

To account for these cases we use the following model atmospheres:
\begin{itemize}
\item For stars with \teff\ $\ge$ 20000 K we constructed
an extensive grid of pure H and He plane parallel model atmospheres
using the non-LTE code {\em TLUSTY} code of Hubeny \& Lanz (1995).
As far as it is possible to construct hydrostatic structures,
we calculated a model grid with various $\log g$ values 
(typically $\Delta \log g = 0.5$) to cover the domain corresponding 
to the evolutionary tracks (see below).
The primordial He abundance of Izotov \& Thuan (1998, mass fraction
$Y=0.244$) was adopted. 
The influence of the exact H/He abundance on the spectra is small.

\item To explore the possible importance of mass loss on the 
ionising spectra of massive Pop III stars, we use the {\em CMFGEN} code
of Hillier \& Miller (1998), which calculates the radiation transfer in the
co-moving frame and solves for non-LTE equilibrium in radiative equilibrium.
The input atmospheric structure including photosphere and wind 
is calculated with the {\em ISA-WIND} code of de Koter \etal\ (1996).
Models were calculated for stellar parameters corresponding to the 
zero age main sequence (ZAMS) of 25, 60, and 500 \msun\ stars 
(parameters given in Sect.\ \ref{s_ave}) and 
an evolved 60 \msun\ model at 52 kK, and wind parameters $v_\infty \sim$ 
2000 km s$^{-1}$ and $\dot{M} =$ 10$^{-9}$ to 10$^{-6}$ \msun\ yr$^{-1}$ 
(10$^{-8}$ to 10$^{-5}$ for the 500 \msun\ model).
The same H/He abundances as above are adopted.

For stars with strong mass loss (as also explored below) we have
used the pure Helium Wolf-Rayet (WR) atmosphere models
of Schmutz \etal\ (1992).
The two parameters, core temperature and wind density,
required to couple these models to the stellar evolution models 
are calculated as in Schaerer \& Vacca (1998).

\item The plane parallel LTE models of Kurucz (1991) with a very metal-poor
composition ([Fe/H] $= -5.$) are adopted for stars with \teff\ $<$ 20000 K.
The models with such a low metal content are equivalent to pure H and He
LTE models (Cojazzi \etal\ 2000, Chavez \etal\ 2001).

\end{itemize}

\begin{figure}
\centerline{\psfig{file=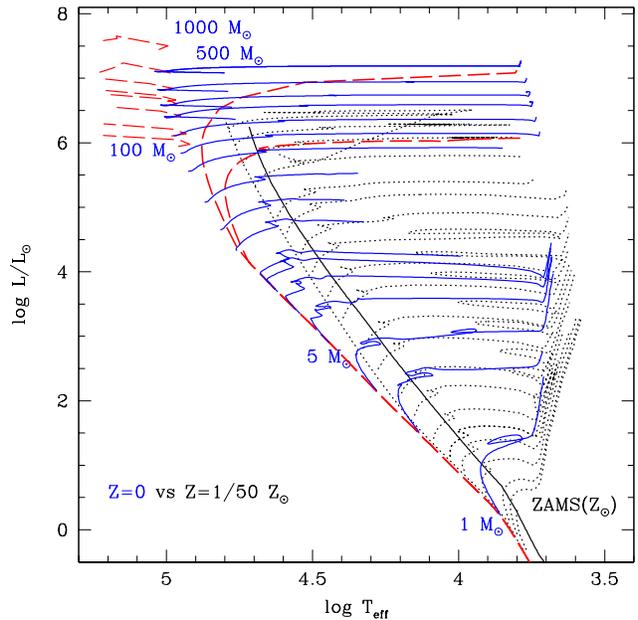,width=8.8cm}}
\caption{HR--diagram for metal free ($Z=0$, solid and long-dashed lines) 
and low metallicity ($Z=1/50 \zsun$, dotted) stars.
Isochrones of 2 and 4 Myr for $Z=0$ tracks without mass loss are also 
plotted (long-dashed). The short dashed high mass tracks evolving blueward
of the ZAMS are computed assuming strong mass loss.
The position of the ZAMS at solar metallicity (Schaller \etal\ 1992) is 
shown by the solid line.
Pop III tracks and isochrones from Marigo \etal\ (2001, no mass loss) and 
Klapp (1983), and El Eid \etal\ (1983, strong mass loss).
$Z=0.0004$ tracks for 0.8 to 150 \msun\ from Lejeune \& Schaerer (2001).
Note the important shift of the ZAMS to high \teff\ from low metallicity
to $Z=0$, as well as the rapid redward evolution of the massive stars}
\label{fig_hrd}
\end{figure}

\subsection{Stellar tracks}
To explore two extreme evolutionary scenarios for massive stars we have
constructed two different sets of zero metallicity stellar evolution tracks:
{\em 1)} no mass loss, {\em 2)} strong mass loss (see Fig.\ \ref{fig_hrd}).

For tracks with no or negligible mass loss (1) we use recent tracks from
1 to 500 \msun\ calculated with the Geneva stellar evolution code (Feij\'oo 
1999, Desjacques 2000). These tracks, including only the core H-burning phase
and assuming small mass loss, have been compared with the tracks of
Marigo \etal\ (2001) up to 100 \msun\ and additional models up to
500 \msun\ by Marigo (2000, private communication).
Good agreement is found regarding the zero age main sequences (ZAMS),
H-burning lifetimes, and the overall appearance of the tracks. 
The differences due to mass loss in the Geneva models are minor for the 
purpose of the present work. 
Furthermore, as the He-burning lifetime is $\la$ 10 \% of
the main sequence phase and is spent at cooler temperatures (cf.\ Marigo \etal\ 
2001), neglecting this phase has no consequences for our predictions.
This has also been verified by comparisons of integrated stellar populations
adopting alternatively the isochrones provided by Marigo \etal\ (2001).

The ``strong mass loss'' set (2) consists of a combination of the following
stellar tracks for the mass range from 80 to 1000 \msun.
Tracks of Klapp (1983) for initial masses of 1000 and 500 \msun\ computed with
values of $N=50$ and 100 respectively for the mass loss parameter\footnote{Tracks 
with these $N$ values yield a qualitatively similar
behaviour as the El Eid \etal\ (1983) models.}.
For 300, 220, 200, 150, 100, and 80 \msun\ we adopt the tracks of El Eid
\etal\ (1983).
The remaining models (1 to 60 \msun) are from the ``no mass loss'' set.
The main difference between the ``strong'' and ``no mass loss'' sets
is the rapid {\em blueward} evolution of the stars in the former case,
due to strong increase of the He abundance on the surface of these stars
leading to a hot WR-like phase (see Fig.\ \ref{fig_hrd}).
While the use of updated input physics (e.g.\ nuclear reaction
rates) could lead to somewhat different results if recomputed with
more modern codes, the predicted tracks depend essentially only
on the adopted mass loss (see also Chiosi \& Maeder 1986), 
and remain thus completely valid.

In Figure \ref{fig_hrd} the two set of tracks are
compared to the low metallicity ($Z=1/50$ \zsun, dotted lines) models 
of Lejeune \& Schaerer (2000) and the position of a solar metallicity
ZAMS (Schaller \etal\ 1992).
This illustrates the well known fact that -- due to the lack of 
CNO elements -- the ZAMS of massive Pop III stars is much hotter that 
their solar or low metallicity counterparts (cf.\ Ezer \& Cameron 1971, 
El Eid \etal\ 1983, Tumlison \& Shull 2000).
In particular this implies that at $Z=0$ stars with $M \ga$ 5 \msun\ have
unusually high temperatures
and in turn non-negligible ionising fluxes (cf.\ Sect.\ 
\ref{s_single}) corresponding to normal O-type stars (\teff\ $\ga$ 30 kK).

\subsection{Evolutionary synthesis models}
To calculate the properties of integrated zero metallicity stellar populations
we have included the stellar atmosphere models and evolutionary tracks
in the evolutionary synthesis code of Schaerer \& Vacca (1998).
The following changes have been made to adapt the calculations to 
metal-free populations.

\subsubsection{Nebular emission}
The potentially strongest H and He recombination lines in the rest-frame UV 
from \lya\ longward to \ha\ in the optical spectrum are calculated, namely
\lya, \Heiiuv, \heii\ $\lambda$3203, \hei\ $\lambda$4026, \Hei, \Heiiopt,
\hb, \hei\ $\lambda$5016, \hei\ $\lambda$5876, and \ha\footnote{More precisely: 
all \heii\ lines with relative intensities
$I_\lambda/I_{\hb} \ga$ 0.4, \hei\ with $I_\lambda/I_{4471} \ga$ 0.4, plus \hb\ 
and \ha.}.
We assume Case B recombination for an electron temperature of $T_e=30000$ K 
appropriate of metal-free gas and a low electron density ($n_e = 100$ cm$^{-3}$). 
The emissivities and recombination coefficients of Storey \& Hummer (1995) 
are used in general. 
For \Hei\ we take the data from Osterbrock (1989) for  $T_e=20000$ K. 
\lya\ emission is computed assuming a fraction of 0.68 of photons converted
to \lya\ (Spitzer 1978).
For these assumptions the line luminosities (in units of erg s$^{-1}$) are thus 
expressed as
\begin{equation}
  L_l \,\, [{\rm erg \, s^{-1}}] = c_l \,\, (1 - f_{\rm esc}) \,\, Q_i \,\,\, [{\rm s^{-1}}],
  \label{eq_lines}
\end{equation}
where $c_l$ are the coefficients given in Table \ref{tab_lines},
$f_{\rm esc}$ is the photon escape fraction out of the idealised region
considered here (we assume  $f_{\rm esc}=0 $ in our calculations, cf.\ Sect.\
\ref{s_pops}),
and $Q_i$ is the ionising photon flux (in units of s$^{-1}$) corresponding to the 
appropriate recombination line.
The line emissivities depend only relatively weakly on the assumed 
conditions. Adopting, e.g.\ a temperature of $T_e=10000$ K leads to 
an increase of the \heii\ line luminosities by $\sim$ 10 -- 30 \%,
whereas changes of $\la$ 10 \% are found for the H lines
($c_{\rm 1640} = 6.40 \times 10{-12}$, $c_{{\rm H}\alpha} = 1.37 \times 10^{-12}$).
This lower value of $T_e$ is adopted for calculations at non-zero metallicities
as in Schaerer \& Vacca (1998)

\begin{table}
\caption{Line emission coefficients $c_l$ for Case B,  $T_e=30000$ K, 
$n_e = 100$ cm$^{-3}$}
\label{tab_lines}
\begin{tabular}[htb]{lllllll}
Line & $c_l$ [erg] & appropriate ionising flux $Q_i$ \\
\hline \\ \smallskip
\lya\     & $1.04 \times 10^{-11}$ & \Qh \\
\Heiiuv\  & $5.67 \times 10^{-12}$ & \Qhep \\
\heii\ $\lambda$3203 &  $3.32 \times 10^{-13}$ & \Qhep \\
\hei\ $\lambda$4026      & $9.25 \times 10^{-13}$ & \Qhe \\     
\Hei\      & $1.90 \times 10^{-13}$ & \Qhe \\
\Heiiopt\ & $7.13 \times 10^{-13}$ & \Qhep \\
\hb\      & $4.47 \times 10^{-13}$ & \Qh \\
\hei\ $\lambda$5016      & $1.31 \times 10^{-13}$ & \Qhe \\     
\hei\ $\lambda$5876      & $4.90 \times 10^{-13}$ & \Qhe \\
\ha\      & $1.21 \times 10^{-12}$ & \Qh \\
\smallskip \\ \hline 
\end{tabular}
\end{table}

As we will be studying objects with strong ionising fluxes,
nebular continuous emission must also be included.
The monochromatic luminosity of the gas is given by
\begin{equation}
  \label{eq_lcont}
  L_\lambda = \frac{c}{\lambda^2} \frac{\gamma_{\rm total}}{\alpha_B} (1 - f_{\rm esc}) Q({\rm H})
\end{equation}
(e.g.\ Osterbrock 1989),
where $\alpha_B$ is the case B recombination coefficient for hydrogen.
The continuous emission coefficient $\gamma_{\rm total}$, including
free-free and free-bound emission by H, neutral He and 
singly ionised He, as well as the two-photon continuum of hydrogen is given
by
\begin{equation}
  \label{eq_gamma}
  \gamma_{\rm total} = \gamma_{\rm HI} + \gamma_{\rm 2q}
        +  \gamma_{\rm HeI} \frac{n({\rm He}^+)}{n({\rm H}^+)}
        +  \gamma_{\rm HeII} \frac{n({\rm He}^{++})}{n({\rm H}^+)}.
\end{equation}
The emission coefficients $\gamma_i$ 
(in units of erg cm$^3$ s$^{-1}$ Hz$^{-1}$)
are taken from the tables of
Aller (1984) and Ferland (1980) for wavelength below/above 1 $\mu$m 
respectively for an electron temperature of 20000 K
\footnote{For non-zero metallicities we adopt again $T_e$ = 10000 K
as in Schaerer \& Vacca (1998).}.
The primordial H/He abundance ratio of Izotov \& Thuan (1998) is adopted.
As a non-negligible \heii\ ionising flux is obtained in the calculations
under certain conditions (cf.\ below), we also allow for possible 
continuous emission from the region with doubly ionised He.
The relative contribution of the He$^{++}$ and the He$^+$ regions
is to first order easily calculated from the relative size of the 
respective Stroemgren spheres and is given by 
$(n({\rm He}^{++})/n({\rm H}^+)) / (n({\rm He}^+)/n({\rm H}^+)) 
 \sim 2. \times Q({\rm He^+})/Q({\rm H})$.
As expected, the resulting contribution from the He$^+$ region is small,
even in the case of the hardest spectra considered here.
However, as shown later, the total nebular continuous emission is not negligible for
Population III objects even in the rest-UV range.

\subsubsection{Initial mass function}
The question of the masses and mass distribution of metal-free stars
has been studied since the 1960's (see the volume of Weiss \etal\ 2000
for recent results).

While the outcome from various hydrodynamical models and other studies
still differ, there seems to be an overall consensus that
stars with unusually large masses (up to $\sim$ 10$^3$ \msun) 
may form, possibly even preferentially (e.g.\ Abel \etal\ 1998, Bromm \etal\ 
1999, 2001, Nakamura \& Umemura 2001).
Uehara \etal\ (1996) and Nakamura \& Umemura (1999, 2001) also find 
that the formation of stars with masses down to $\sim$ 1 \msun\ is not 
excluded (cf.\ also Abel \etal\ 2001). 
The differences may be of various origins (adopted numerical 
scheme and resolution, treatment of radiation transfer and optically thick
regions etc.).

In view of our ignorance on this issue we adopt a variety of different
upper and lower mass limits of the IMF assumed to be a powerlaw, with the aim of 
assessing their impact on the properties of integrated stellar populations.
The main cases modeled here are summarised below in Table \ref{tab_models}.
Except if mentioned otherwise, the IMF slope is taken as the 
Salpeter value ($\alpha =$ 2.35) between the lower and upper 
mass cut-off values \mlow\ and \mup\ respectively.
The consideration of stars even more massive than 1000 \msun\ (or no mass loss
models with $>$ 500 \msun) is limited by the availability of evolutionary 
tracks for such objects.
The tabular data given below allows the calculation of integrated populations
with arbitrary IMFs for ZAMS populations and the case of a constant SFR .

\begin{table*}
\caption{
Summary of main model input parameters. All models assume a Salpeter
slope for the IMF}
\label{tab_models}
\begin{tabular}[htb]{llrrr}
Model ID & tracks & \mlow\ & \mup\ & symbol in Figs.\
  \protect\ref{fig_i},  \protect\ref{fig_w1}, \protect\ref{fig_w2} \\ \smallskip 
\\ \hline
A         & No mass loss &  1   & 100  &  dotted \\
B         & No mass loss &  1   & 500  &  solid \\
C         & No mass loss &  50  & 500  &  long dashed \\
\smallskip \\
D     & Strong mass loss &  1   & 500  & dash -- dotted \\
E     & Strong mass loss  & 50  & 1000 & long dash -- dotted \\
\smallskip \\
 \multicolumn{5}{l}{Solar metallicity tracks:}   \\
ZS       & High mass loss & 1   & 100  \\
 \multicolumn{5}{l}{$Z=1/50$ \zsun\ tracks:}   \\
ZL       & Mass loss      & 1   & 150  \\
 \hline
\end{tabular}
\end{table*}

\section{Properties of individual stars}
\label{s_single}

\begin{figure}
\centerline{\psfig{file=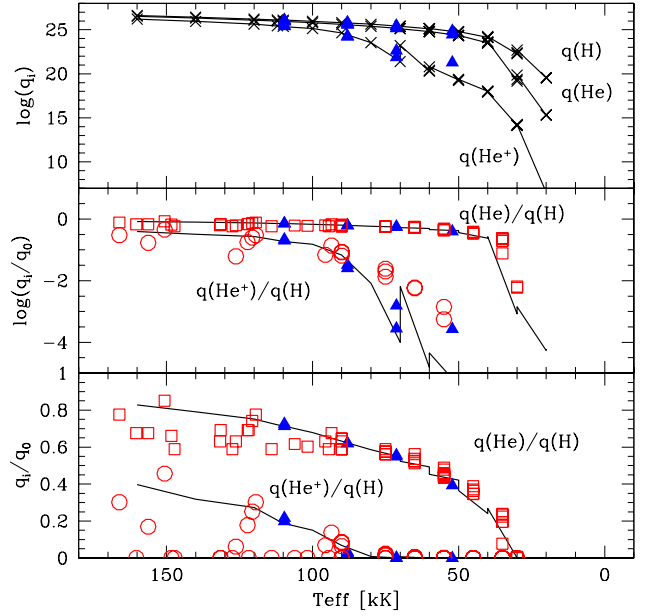,width=8.8cm}}
\caption{Ionising photon flux (top panel) and hardness of the ionising 
spectrum (middle and bottom panel) as a function of effective temperature 
for various atmosphere models.
Solid lines connect plane parallel {\em TLUSTY} Pop III composition models 
of various \teff\ and $\log g$. Triangles show calculations from the spherically 
expanding {\em CMFGEN} models for Pop III stars. 
For comparison the pure H and He WR models of Schmutz \etal\ (1992) are also 
shown (open squares and circles).
Discussion in text}
\label{fig_qi_teff}
\end{figure}

\begin{figure}
\centerline{\psfig{file=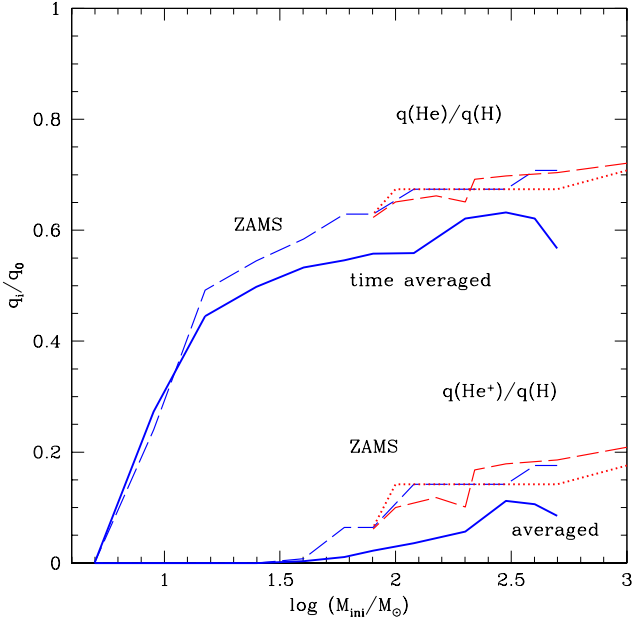,width=8.8cm}}
\caption{
Hardness of the ionising spectrum expressed by \qrathe\ and \qrathep\
as a function of the initial stellar mass.
Dashed lines show the hardness of all ZAMS models.
The thick lines (dashed: no mass loss, dotted: high mass loss tracks)
show the average spectral hardness when stellar evolution effects are 
taken into account.
Note in particular the importance of stellar evolution leading to a decrease 
of the hardness of the \heii\ ionising spectrum by approximately a factor of 2
}
\label{fig_qi_ave}
\end{figure}

\begin{table*}[htb]
\caption{Properties of Pop III ZAMS stars}
\begin{tabular}{rrrrrrrrrr}
$M_{\rm ini}$ & $\log L/\lsun$ & $\log \teff$ & \Qh & \Qhe & \Qhep
  & \Qhtwo & \Qrathe & \Qrathep \\
\hline \\ 
1000. & 7.444 & 5.026 & 1.607E+51 &1.137E+51 &2.829E+50 &1.727E+51 &0.708E+00 &0.176E+00\\
 500. & 7.106 & 5.029 & 7.380E+50 &5.223E+50 &1.299E+50 &7.933E+50 &0.708E+00 &0.176E+00\\
 400. & 6.984 & 5.028 & 5.573E+50 &3.944E+50 &9.808E+49 &5.990E+50 &0.708E+00 &0.176E+00\\
 300. & 6.819 & 5.007 & 4.029E+50 &2.717E+50 &5.740E+49 &4.373E+50 &0.674E+00 &0.142E+00\\
 200. & 6.574 & 4.999 & 2.292E+50 &1.546E+50 &3.265E+49 &2.487E+50 &0.674E+00 &0.142E+00\\
 120. & 6.243 & 4.981 & 1.069E+50 &7.213E+49 &1.524E+49 &1.161E+50 &0.674E+00 &0.142E+00\\
  80. & 5.947 & 4.970 & 5.938E+49 &3.737E+49 &3.826E+48 &6.565E+49 &0.629E+00 &0.644E-01\\
  60. & 5.715 & 4.943 & 3.481E+49 &2.190E+49 &2.243E+48 &3.848E+49 &0.629E+00 &0.644E-01\\
  40. & 5.420 & 4.900 & 1.873E+49 &1.093E+49 &1.442E+47 &2.123E+49 &0.584E+00 &0.770E-02\\
  25. & 4.890 & 4.850 & 5.446E+48 &2.966E+48 &5.063E+44 &6.419E+48 &0.545E+00 &0.930E-04\\
  15. & 4.324 & 4.759 & 1.398E+48 &6.878E+47 &2.037E+43 &1.760E+48 &0.492E+00 &0.146E-04\\
   9. & 3.709 & 4.622 & 1.794E+47 &4.303E+46 &1.301E+41 &3.785E+47 &0.240E+00 &0.725E-06\\
   5. & 2.870 & 4.440 & 1.097E+45 &8.629E+41 &7.605E+36 &3.760E+46 &0.787E-03 &0.693E-08\\
\hline \\
\end{tabular}
\label{tab_zams}
\end{table*}

\begin{table*}[htb]
\caption{Time averaged properties calculated from Pop III tracks with no mass loss}
\begin{tabular}{rrrrrrrrrr}
$M_{\rm ini}$ & lifetime & \Qhave & \Qheave & \Qhepave & \Qhtwoave & \Qratheave & \Qrathepave \\
\hline \\ 
   1000. & \multicolumn{4}{c}{not available} \\
    500.00 &1.899E+06 & 6.802E+50 &3.858E+50 &5.793E+49 &7.811E+50 &0.567E+00 &0.852E-01\\
    400.00 &1.974E+06 & 5.247E+50 &3.260E+50 &5.567E+49 &5.865E+50 &0.621E+00 &0.106E+00\\
    300.00 &2.047E+06 & 3.754E+50 &2.372E+50 &4.190E+49 &4.182E+50 &0.632E+00 &0.112E+00\\
    200.00 &2.204E+06 & 2.624E+50 &1.628E+50 &1.487E+49 &2.918E+50 &0.621E+00 &0.567E-01\\
    120.00 &2.521E+06 & 1.391E+50 &7.772E+49 &5.009E+48 &1.608E+50 &0.559E+00 &0.360E-01\\
     80.00 &3.012E+06 & 7.730E+49 &4.317E+49 &1.741E+48 &8.889E+49 &0.558E+00 &0.225E-01\\
     60.00 &3.464E+06 & 4.795E+49 &2.617E+49 &5.136E+47 &5.570E+49 &0.546E+00 &0.107E-01\\
     40.00 &3.864E+06 & 2.469E+49 &1.316E+49 &8.798E+46 &2.903E+49 &0.533E+00 &0.356E-02\\
     25.00 &6.459E+06 & 7.583E+48 &3.779E+48 &3.643E+44 &9.387E+48 &0.498E+00 &0.480E-04\\
     15.00 &1.040E+07 & 1.861E+48 &8.289E+47 &1.527E+43 &2.526E+48 &0.445E+00 &0.820E-05\\
      9.00 &2.022E+07 & 2.807E+47 &7.662E+46 &3.550E+41 &5.576E+47 &0.273E+00 &0.126E-05\\
      5.00 &6.190E+07 & 1.848E+45 &1.461E+42 &1.270E+37 &6.281E+46 &0.791E-03 &0.687E-08\\
\hline \\                                                                            
\end{tabular}
\label{tab_ave}
\end{table*}

\begin{table*}[htb]
\caption{Time averaged properties calculated from Pop III tracks with strong mass loss}
\begin{tabular}{rrrrrrrrrr}
$M_{\rm ini}$ & lifetime & \Qhave & \Qheave & \Qhepave & \Qhtwoave & \Qratheave & \Qrathepave \\
\hline \\ 
   1000. &2.430E+06 & 1.863E+51 &1.342E+51 &3.896E+50 &2.013E+51 &0.721E+00 &0.209E+00\\
    500. &2.450E+06 & 7.719E+50 &5.431E+50 &1.433E+50 &8.345E+50 &0.704E+00 &0.186E+00\\
    300. &2.152E+06 & 4.299E+50 &3.002E+50 &7.679E+49 &4.766E+50 &0.698E+00 &0.179E+00\\
    220. &2.624E+06 & 2.835E+50 &1.961E+50 &4.755E+49 &3.138E+50 &0.692E+00 &0.168E+00\\
    200. &2.628E+06 & 2.745E+50 &1.788E+50 &2.766E+49 &3.028E+50 &0.651E+00 &0.101E+00\\
    150. &2.947E+06 & 1.747E+50 &1.156E+50 &2.066E+49 &1.917E+50 &0.662E+00 &0.118E+00\\
    100. &3.392E+06 & 9.398E+49 &6.118E+49 &9.434E+48 &1.036E+50 &0.651E+00 &0.100E+00\\
     80. &3.722E+06 & 6.673E+49 &4.155E+49 &4.095E+48 &7.466E+49 &0.623E+00 &0.614E-01\\
\hline \\                                                                               
\end{tabular}                                                                           
\label{tab_ave_mdot}                                                                    
\end{table*}                                                                            

\begin{table}[tb]
\caption{Least square fits to the time averaged properties for Pop III stars,
metal-poor objects ($Z = 1/50$ \zsun) and solar metallicity stars for comparison.
The time averaged quantities  
$\log y(M) = a_0 + a_1 x + a_2 x^2 + a_3 x^3$ (col.\ 1)
are fitted as a function of the initial mass ($x=\log (M/\msun)$) 
as 2nd or 3rd order polynomials over the mass interval given in col.\ 2. 
For masses lower than the range considered the corresponding photon
flux $\bar{Q}$ is negligible.
$\bar{Q}$ values are in units of photon $s^{-1}$, stellar lifetimes
$t_\star$ in yr.
}
\begin{tabular}{llrrrrrrrr}
Quantity $y$ & mass range & $a_0$ & $a_1$ & $a_2$ & $a_3$\\
\hline \\ 
 \multicolumn{3}{l}{$Z=0$ tracks with no mass loss:}   \\
\Qhave   & 9 -- 500 \msun & 43.61 & 4.90 & -0.83 \\  
\Qhave   & 5 -- 9 \msun   & 39.29 & 8.55 &       \\  
\Qheave  & 9 -- 500 \msun & 42.51 & 5.69 & -1.01 \\  
\Qheave  & 5 -- 9 \msun   & 29.24 & 18.49&       \\  
\Qhepave & 5 -- 500 \msun & 26.71 & 18.14& -3.58 \\  
\Qhtwoave& 5 -- 500 \msun & 44.03 & 4.59 & -0.77 \\  
$t_\star$ &5 -- 500 \msun & 9.785 & -3.759 & 1.413 & -0.186 \\  
 \multicolumn{3}{l}{$Z=0$ tracks with strong mass loss:} \\
\Qhave   & 80 -- 1000 \msun & 46.21 & 2.29 & -0.20 \\    
\Qheave  & 80 -- 1000 \msun & 45.71 & 2.51 & -0.24 \\    
\Qhepave & 80 -- 1000 \msun & 41.73 & 4.86 & -0.64 \\    
\Qhtwoave& 80 -- 1000 \msun & 46.25 & 2.30 & -0.21 \\    
$t_\star$ & 80 -- 1000 \msun& 8.795 & -1.797 & 0.332 \\  
 \multicolumn{3}{l}{Solar metallicity tracks:}   \\
\Qhave   & 7 -- 120 \msun & 27.89 & 27.75 & -11.87 & 1.73 \\
\Qheave  & 7 -- 120 \msun & 1.31  & 64.60 & -28.85 & 4.38 \\
\Qhtwoave& 15 -- 120 \msun& 41.90 & 7.10  & -1.57  \\
$t_\star$ & 7 -- 120 \msun& 9.986 & -3.497 & 0.894 \\
 \multicolumn{3}{l}{$Z=1/50$ \zsun\ tracks:}   \\
\Qhave   & 7 -- 150 \msun & 27.80 & 30.68 & -14.80 & 2.50 \\
\Qheave  & 20 -- 150 \msun& 16.05 & 48.87 & -24.70 & 4.29 \\
\Qhepave & 20 -- 150 \msun& 34.65 & 8.99  & -1.40 \\
\Qhtwoave& 12 -- 150 \msun& 43.06 & 5.67  & -1.08 \\
$t_\star$& 7 -- 150 \msun & 9.59  & -2.79 &  0.63 \\
\hline \\                                                                            
\end{tabular}
\label{tab_fits}
\end{table}

We now discuss the properties of the ionising spectra and their dependence
on assumptions of model atmospheres.
The ZAMS properties of individual stars as well as their average properties
taken over their lifetime (i.e.\ including the effect of stellar evolution)
are presented next.

\subsection{Ionising and  H$_2$ dissociating spectra}
\label{s_single_zams}
The ionising photon fluxes $q_i$ and the hardness of the spectrum
(described by \qrathe\ and \qrathep) predicted by various atmosphere
models as a function of \teff\ are shown in Fig.\ \ref{fig_qi_teff}.
The ionising photon flux per surface area $q_i$, given in units of 
photons s$^{-1}$ cm$^{-2}$, is related to the stellar ionising photon
flux $Q_i$ (in photons s$^{-1}$) used later by
\begin{equation}
  \label{eq_qi}
  Q_i = 4 \pi R_\star^2 q_i = 4 \pi R_\star^2 \int_{\nu_i}^{\infty} \frac{F_\nu}{h \nu} d\nu,
\end{equation}
where $R_\star$ is the stellar radius, $F_\nu$ the spectral flux distribution
(in erg cm$^{-2}$ s$^{-1}$ Hz$^{-1}$), and $\nu_i$ the frequency of the 
appropriate ionisation potential.
The photon flux \Qhtwo\ in the Lyman-Werner band (11.2 to 13.6 eV) 
capable to dissociate H$_2$ is also calculated.

The solid lines in Fig.\ \ref{fig_qi_teff} give the predictions obtained 
from the plane parallel non-LTE {\em TLUSTY} models.
As already discussed by Tumlinson \& Shull (2000, TS00), 
this shows in particular the high fraction of photons emitted with energies 
capable to ionise \hei\ and \heii\ for stars with \teff\ $\ga$ 40 kK and 
$\ga$ 80 kK respectively.

As the stellar tracks with strong mass loss evolve from the ZAMS (at
$\sim$ 100 kK for $M \ga 100 \msun$) blueward, we have also extended the 
temperature range to the hottest temperatures for which atmosphere
models with winds are available (Schmutz \etal\ 1992, shown as open symbols
in Fig.\ \ref{fig_qi_teff}).
As best shown in the bottom panel, the ratios \qrathe\ and \qrathep\ 
continue to increase with \teff, as the maximum of the flux distribution
has not yet shifted above 54 eV. 
However, it is important to remember that in the case of atmospheres with
winds, the ionising flux also depends on the wind density as illustrated
by the dispersion of the Schmutz \etal\ models for given \teff.
For example, for sufficiently dense winds the \heii\ ionising flux can
be completed suppressed (cf.\ Schmutz \etal\ 1992).
However, for the wind densities obtained in the present models, this situation
is not encountered (cf.\ below).

The ionising properties of ZAMS models for masses between 5 and 1000 \msun\ are
given in Table \ref{tab_zams}\footnote{The luminosities and \teff\ are taken
from the set of tracks without mass loss, complemented by the 1000 \msun\ model
of Klapp (1983). As expected the differences between the various ZAMS models
is small. The approximate stellar models of Bromm \etal\ (2001) and TS00
yield somewhat higher \teff\ by up to $\sim$ 6 kK and 30 kK respectively.}.

The following differences are obtained between various atmosphere models
for \teff\ between 50 and 100 kK.
The plane parallel {\em TLUSTY} models show some variation of \qhep\ at 70 
kK with gravity, with low $\log g $ leading to higher \qhep.
As expected, for very hot stars the {\em CMFGEN} photosphere-wind models 
(filled triangles in Fig.\ \ref{fig_qi_teff}) show little deviation from 
the plane parallel models, since the \heii\ continuum is formed in progressively
deeper layers of the atmosphere (e.g.\ Husfeld \etal\ 1984, Clegg \& Middlemass 1987).
The two models at \teff\ $\sim$ 70 kK (corresponding to a ZAMS mass of 25 \msun)
show an increase of \qhep\ of up to $\sim$ 1 dex with respect to the 
{\em TLUSTY} model of same $\log g$, due to a depopulation of the \heii\
groundstate induced by the stellar wind (Gabler \etal\ 1989).
Towards lower \teff\ differences between static and spherically expanding
atmospheres becomes progressively more important for \qhep.
However, in the integrated populations considered later, the emission of
\heii\ ionising photons will be dominated by the hottest objects.
We are therefore likely left with the situations where either possible wind 
effects have no significant impact on the total \qrathep\ hardness, or 
\qrathep\ is too small to have observable consequences.
In conclusion, it appears safe to rely only on non-LTE plane parallel 
atmospheres to properly describe the ionising fluxes of metal-free 
stellar populations.

Can other effects, not included in the present model atmosphere grids,
affect the ionising spectra ?
The effect of non-coherent electron scattering could be important for 
\teff\ $\ga$ 40 kK (Rybicki \& Hummer 1994).
Test calculations with the {\em CMFGEN} code of Hillier \& Miller (1998) for 
some of the above {\em CMFGEN} models indicate negligible changes on the 
ionising photon fluxes.
Compton scattering, instead of the commonly implemented Thomson scattering, 
should not affect the ionising fluxes of normal Pop III stars, as differences 
appear only above \teff\ $\ga$ 150 kK (Hubeny \etal\ 2001).
X-rays, originating from stellar wind instabilities or interactions with
magnetic fields, can potentially increase the high energy (He$^+$) ionising
flux in stars with weak winds (MacFarlane \etal\ 1994). 
For Pop III stars such a hypothesis remains highly speculative.
The adopted model atmospheres should thus well describe the spectra
of the objects considered here.

\subsubsection{Comparisons with other Pop III calculations}
\label{s_comp}
As expected, our calculations for individual stars of a given \teff\
agree quite well with those of TS00 based also on the {\em TLUSTY} 
model atmospheres.
As briefly mentioned above, small differences are found with their 
ZAMS positions calculated with simplified stellar models.
For integrated populations the use of a proper ZAMS 
leads to somewhat softer spectra (cf.\ Sect.\ \ref{s_time}).

As pointed out by Bromm \etal\ (2001) the spectral properties of stars
with $M \ga$ 300 \msun\ are essentially independent of stellar mass, 
when normalised to unit stellar mass.
The degree to which this holds can be verified from Table \ref{tab_zams}.
If correct, it would imply that the total spectrum of such a population 
would be independent of the IMF and only depend on its total mass. 
Our results for the 1000 \msun\ ZAMS star are in rather good agreement with 
the ionising fluxes of Bromm \etal\ (2001, their Table 1)
\footnote{Our values for \Qh, \Qhe, and \Qhep\ differ by 0, +3, and - 25 \%
respectively from the values of Bromm \etal.}.
However, as already apparent from their Fig.\ 3 the deviations from this 
approximate behaviour are not negligible for the He$^+$ ionising flux.
In consequence we find e.g.\ \Qh\ reduced by 12 \% and 
\heii\ recombination line strengths reduced by a factor 2 for 
a population with a Salpeter IMF from 300 to 1000 \msun!
Larger difference are obviously obtained for IMFs with \mlow\ $<$ 300 \msun.
Further differences (due to the inclusion of nebular continuous emission)
with Bromm \etal\ (2001) are discussed in Sect.\ \ref{s_pops}.

\begin{figure}[h]
\centerline{\psfig{file=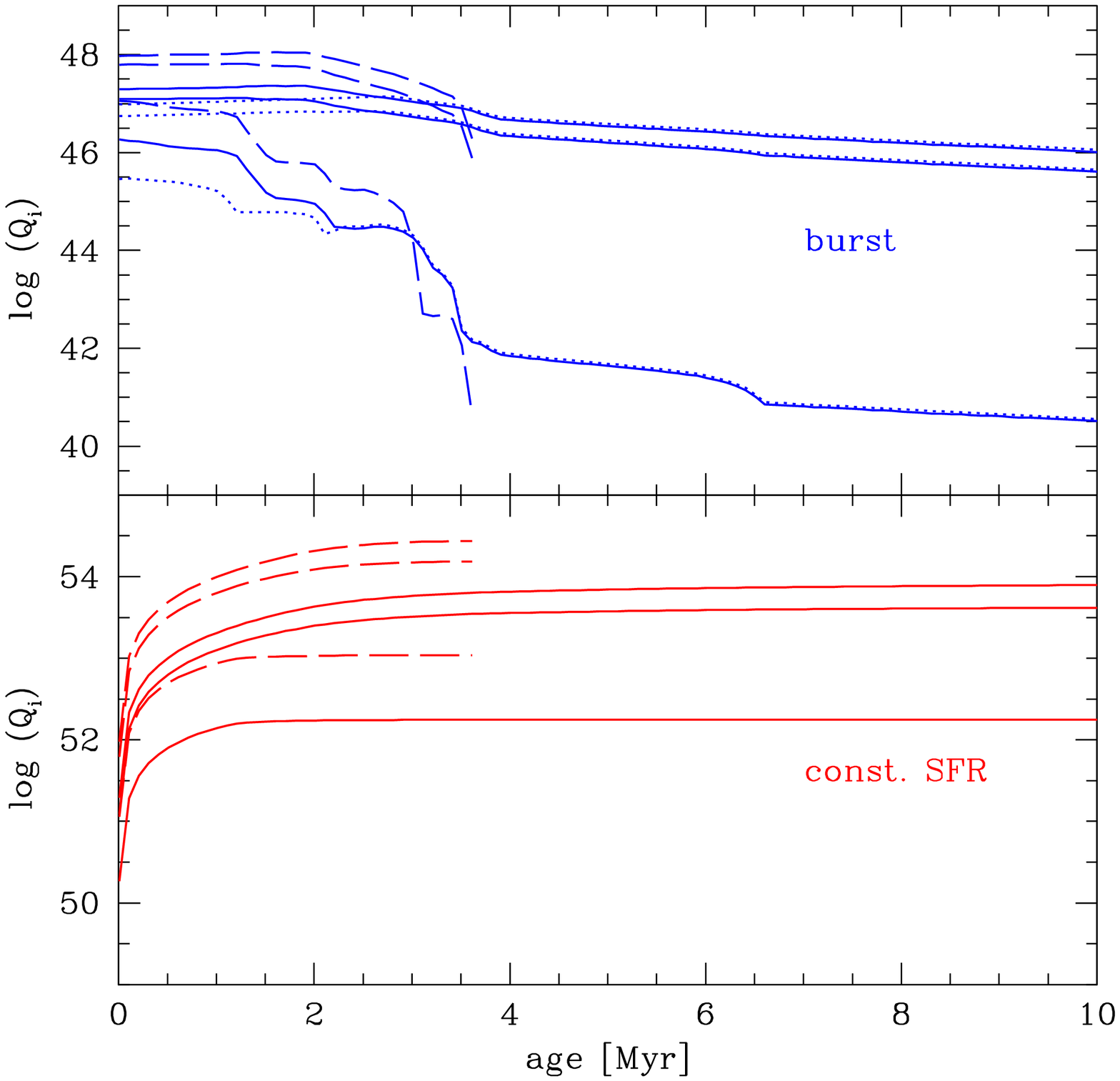,width=8.8cm}}
\caption{
Ionising photon fluxes \Qh, \Qhe\ and \Qhep\ as a function of 
time for instantaneous burst models (upper panel) or constant star formation
(lower panel). For each group of 3 lines (using the same symbols) \Qh, \Qhe\ 
and \Qhep\ are given from top to bottom. 
Different IMFs are illustrated by the dotted (model ID A), solid (model
B) and long-dashed lines (model C) respectively (cf.\ Table \ref{tab_models}).
The units are photon s$^{-1}$ \msun$^{-1}$ for the burst and
photon s$^{-1}$ normalised per unit SFR (in \msun\ yr$^{-1}$) for constant SF.
Note in particular the rapid decrease of the \heii\ ionising flux 
$\sim$ 3--4 Myr after the burst
}
\label{fig_qi}
\end{figure}


\begin{figure*}[htb]
\centerline{\psfig{file=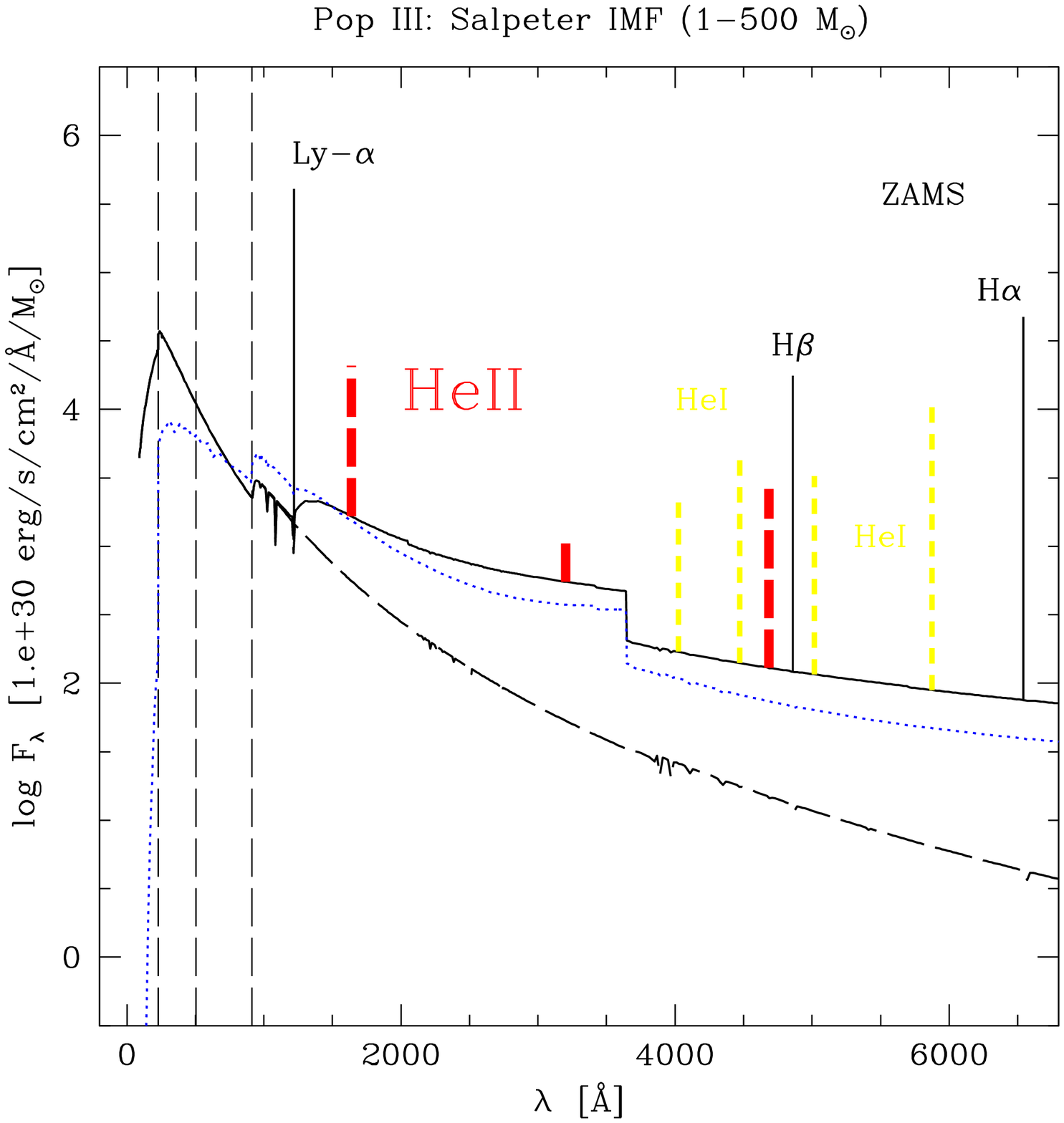,width=8.8cm}
  \psfig{file=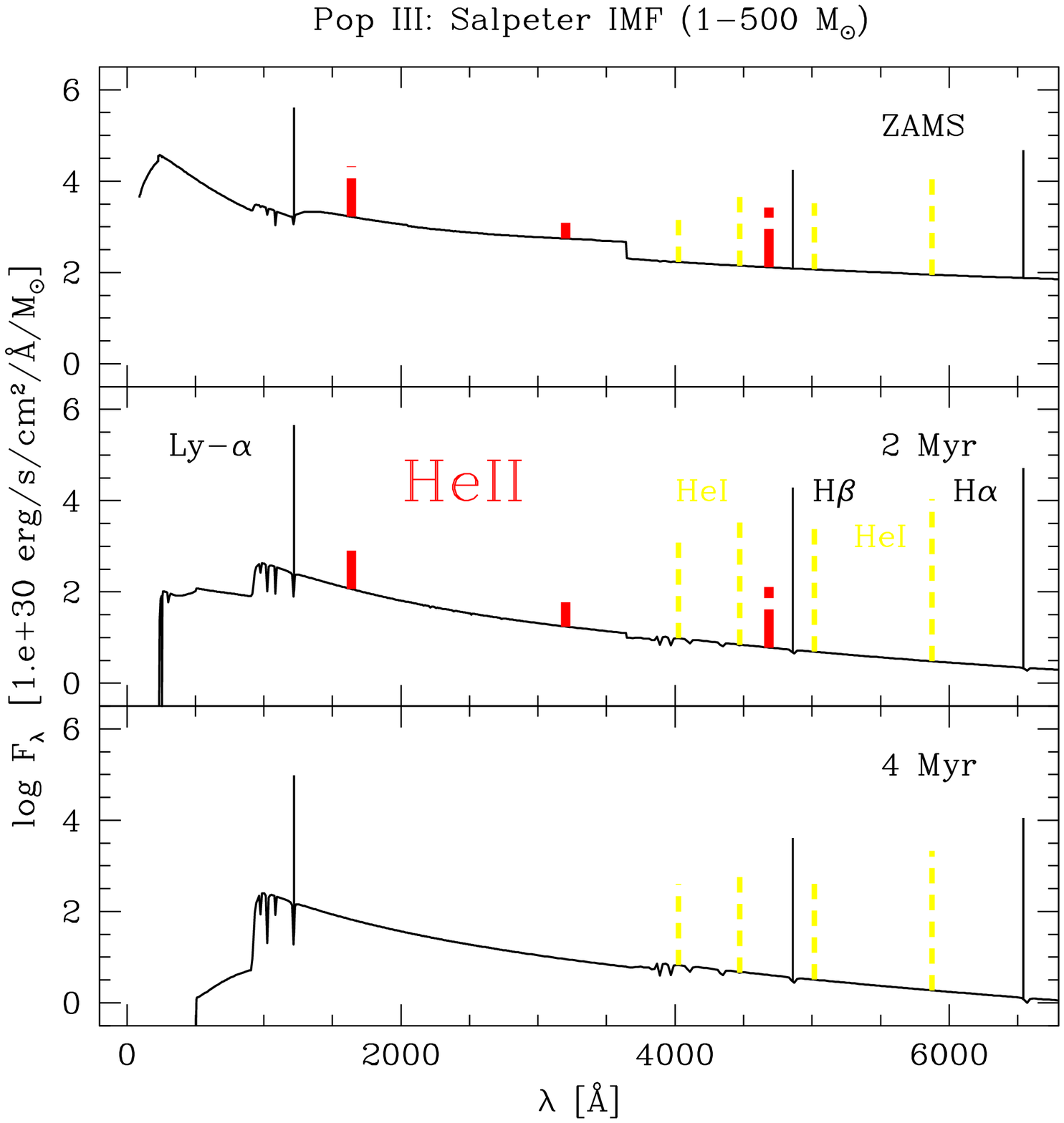,width=8.8cm}}
\caption{Spectral energy distribution (SED) including H and He recombination lines
for model B (solid line, see Table \protect\ref{tab_lines} for line identifications).
{\bf Left panel:} ZAMS population.
The pure stellar continuum (neglecting nebular emission) is shown by the
dashed line. For comparison the SED of the $Z=1/50 \zsun$ population
(model ZL: Salpeter IMF from 1 -- 150 \msun) is shown by the dotted line.
The vertical dashed lines indicate the ionisation potentials of H, He$^0$, and 
He$^+$. 
Note the presence of the unique \heii\ features (shown as thick dashed lines)
and the importance of nebular continuous emission.
{\bf Right panel:} temporal evolution of the spectrum for 0, 2 and 4 Myr
showing the rapid change of the emission line spectrum, characterised by the 
disappearance of the \heii\ lines
}
\label{fig_seds}
\end{figure*}

\subsection{Time averaged ionising properties}
\label{s_ave}

As clear from Fig.\ \ref{fig_hrd}, massive stars evolve rapidly toward
cooler temperatures during their (short) lifetime in the absence of strong 
mass loss. It is thus evident that especially the 
\heii\ ionising flux will strongly decrease with age.
To quantify this effect we have calculated the lifetime averaged ionising
flux $\bar{Q}_i$ along the evolutionary tracks, defined by 
\begin{equation}
  \label{eq_qave}
  \bar{Q}_i(M) = \frac{\int_0^{t_\star(M)} Q_i(t,M) dt}{t_\star(M)},
\end{equation}
where $t_\star$ is the stellar lifetime.
The results calculated for both sets of tracks (no mass loss, strong mass loss)
are given in Tables \ref{tab_ave} and \ref{tab_ave_mdot} respectively.
For convenience we also provide simple fits (linear or quadratic) of
these values as a function of the initial mass $M$ in Table \ref{tab_fits}.
Together with $t_\star$ the average ionising flux  $\bar{Q}_i$ defined in this 
manner in particular 
allows one to calculate the correct values of the ionising flux, recombination
line luminosities etc.\ for integrated populations at equilibrium, i.e.\ for 
the asymptotic value obtained for a constant star formation rate.

The time averaged hardness as a function of mass is compared to
the ZAMS value in Fig.\ \ref{fig_qi_ave}.
For tracks without mass loss, \Qrathepave\ is found to be 
a factor of two or more below the ZAMS value.
Smaller differences are obtained for obvious reasons for \Qratheave.
Only in the case of very strong mass loss, the equilibrium value of
the hardness of the ionising spectrum is found to be essentially 
identical as the ZAMS value.

For Pop I stars the effect of time dependent and time averaged ionising 
fluxes and the spectral hardness based on recent atmosphere models have 
been studied earlier in various contexts (analysis of O star populations 
from integrated spectra, studies  of the diffuse ionised gas; see e.g.\ 
Schaerer 1996, 1998).
For comparison sake the resulting fits for solar metallicity and 
1/50 \zsun\ are also given in Table \ref{tab_fits}.

\section{Population III ``galaxies''}
\label{s_pops}

\subsection{Time evolution of integrated spectra}
\label{s_time}
The temporal evolution of the integrated ionising photon flux 
of an instantaneous burst and the case of constant star formation
is plotted in Fig.\ \ref{fig_qi} for different IMFs.
Most notable is the rapid decrease of the \heii\ ionising flux
in a burst,
due to the evolution of the most massive stars off the hot 
ZAMS and due to their short lifetime.
 
Our values compare as follows with other calculations of \Qh\ available
in the literature:
when scaled to the same lower mass cut-off of the IMF ($\mlow = 0.024$ \msun), 
our ZAMS model for IMF A (Table \ref{tab_models}) agrees with \Qh\ (age $\sim$ 0) 
of Cojazzi \etal\ (2000). The same holds for \Qh\ of Ciardi \etal\ (2000),
when adopting their IMF (Salpeter from 1 to 40 \msun). 
A good agreement is found with \Qh\ of TS00 (ZAMS population with a Salpeter IMF 
from 0.1 to 100 \msun), whereas the slightly cooler ZAMS (cf.\ Sect.\ \ref{s_comp}) 
implies a somewhat softer spectrum (\Qhep/\Qh\ $\sim$ 0.03 instead of 0.05).
Differences with Bromm \etal\ (2001) have already been discussed above
(Sect.\ \ref{s_comp}).

Spectral energy distributions of integrated zero metallicity stellar
populations are shown in Fig.\ \ref{fig_seds} for the case of a Salpeter
IMF from 1 to 500 \msun\ and instantaneous bursts of ages 0 (ZAMS), 2,
and 4 Myr. Overplotted on the continuum (including stellar $+$ nebular 
emission: solid lines) are the strongest emission lines for illustration 
purpose.
In the left panel we show for comparison the spectrum of a burst at low metallicity 
(1/50 \zsun, Salpeter IMF from 1 -- 150 \msun; dotted line).
The striking differences, most importantly in the ionising flux above 
the \heii\ edge ($>$ 54 eV), have already been discussed by TS00.
The comparison of the total spectrum (solid line) with the pure stellar emission
(dashed) illustrates the importance of nebular continuous emission neglected
in earlier studies (TS00, Bromm \etal\ 2001), which dominates the ZAMS 
spectrum at $\lambda \ga$ 1400 \AA.
The nebular contribution, whose emission is proportional to \Qh\
(Eq.\ \ref{eq_lcont}), depends rather strongly on the age, IMF, and 
star formation history. For the parameter space explored here (cf.\ Table
\ref{tab_models}), we find that nebular continuous emission is 
not negligible for bursts with ages $\la$ 2 Myr and for constant star formation
models.

\begin{figure}
\centerline{\psfig{file=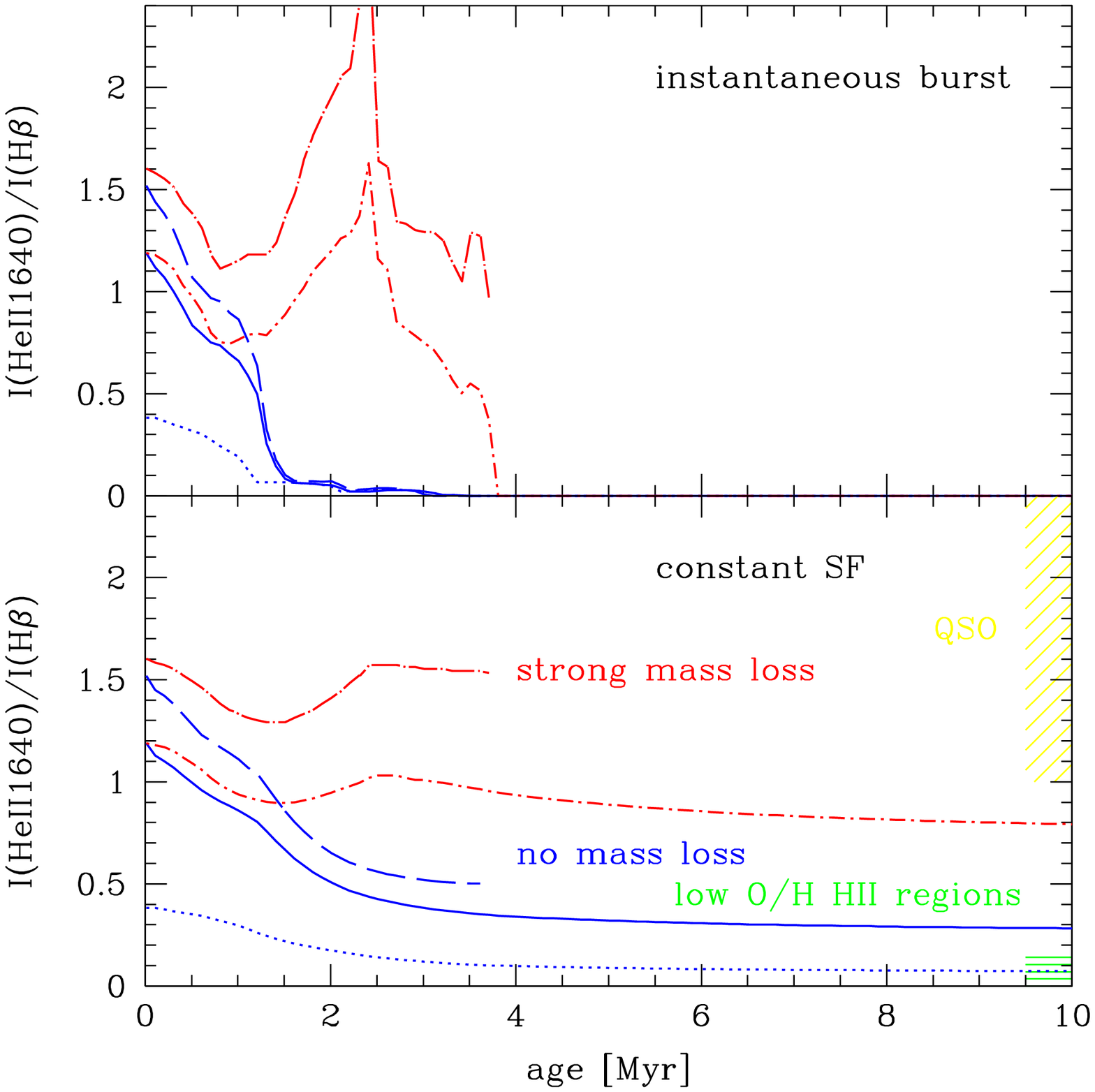,width=8.8cm}}
\caption{Temporal evolution of the relative \Heiiuv/\hb\ line intensity
for various models.
{\bf Upper panel:} instantanous burst models with different IMFs and for 
no mass loss tracks (dotted, dashed, solid) and high mass loss (dash-dotted,
long dash -- dotted; see symbols in Table \protect\ref{tab_models}).
{\bf Lower panel:} same as above for constant star formation models.
The shaded ranges on the right indicate  \Heiiuv/\hb\ intensities
of quasars and maximum values expected for metal-poor \hii\ regions.
For the IMFs considered here $I(1640)/I(\hb)$ can reach the high values 
typical of quasars for very young bursts or in the (extreme) case of 
high mass loss.
}
\label{fig_i}
\end{figure}

The spectra in Fig.\ \ref{fig_seds} show in addition to the H and \hei\ 
recombination lines 
the presence of the 
strong \heii\ $\lambda\lambda$ 1640, 3203, and 4686 recombination lines, which 
--- due to the exceptional hardness of the ionising spectrum ---
represent a unique feature of Pop III starbursts compared to metal enriched 
populations (cf.\ TGS01, Oh \etal\ 2001a, Bromm \etal\ 2001).
Another effect highlighted by this Figure is the rapid temporal evolution of 
the recombination line spectrum. Indeed, already $\ga$ 3 Myr after the burst, 
the high excitation lines are absent, for the reasons discussed before.
In the case of constant star formation, the emission line spectrum at equilibrium is
similar to a burst of $\sim$ 1.3 Myr (cf.\ Fig.\ \ref{fig_i}).

\begin{figure}
\centerline{\psfig{file=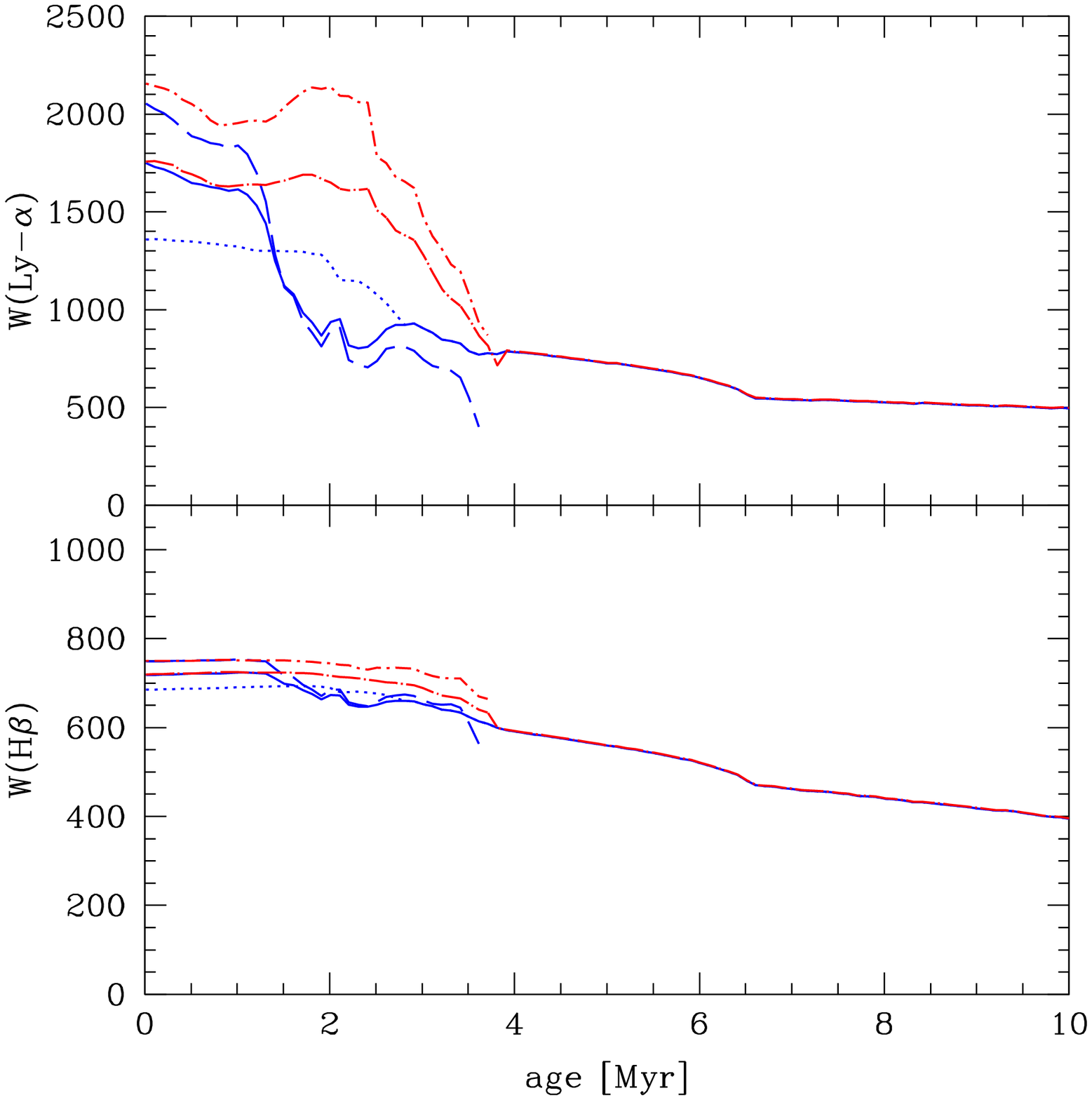,width=8.8cm}}
\caption{Temporal evolution of \lya\ (top) and \hb\ (bottom) for instantaneous burst models.
Note that very large maximum equivalent widths are obtained.
As the IMF and track variations considered here concern only the most
massive stars ($M \protect\ga$ 50 -- 100 \msun) the predictions are only affected
at young ages ($\protect\la$ 4 Myr)
}
\label{fig_w1}
\end{figure}

\begin{figure}
\centerline{
  \psfig{file=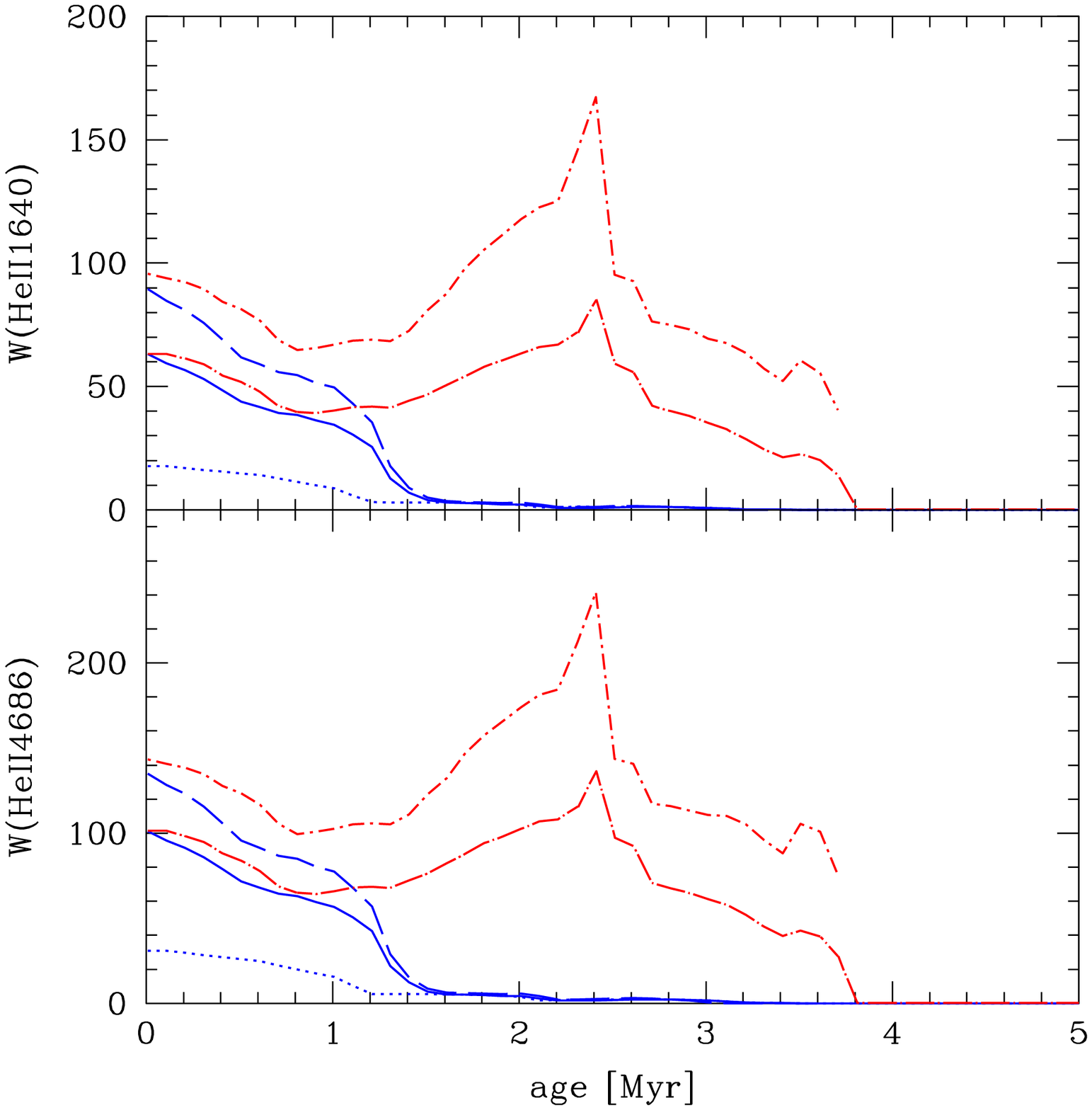,width=8.8cm}}
\caption{Same as \protect\ref{fig_w1} for the \Heiiuv\ (top) and \Heiiopt\ (bottom)
recombination lines. Note that large maximum equivalent widths.}
\label{fig_w2}
\end{figure}

Figure \ref{fig_i} shows the temporal evolution of the line intensity
of \Heiiuv\ with respect to \hb\ for different IMFs, instantaneous
bursts (upper panel) or constant star formation (lower panel),
and the two sets of stellar tracks (see model designations in Table
\ref{tab_models}).
As the lifetime of the stars affected by the IMF variations considered
here is $\sim$ 2 -- 4 Myr, no changes are seen at ages $\ga$ 4 Myr.
The following points are worthwhile noticing:
\begin{itemize}
\item[{\em 1)}] already for a ``normal'' IMF with \mup\ = 100 \msun\ the
maximum \Heiiuv\ intensity is exceptionally large for pure stellar 
photoionisation compared to metal-poor stellar populations (cf.\ TS00);
\item[{\em 2)}] extending the IMF up to 500 -- 1000 \msun\ leads to 
$I(1640)/I(\hb)$ up to $\sim$ 1.5 -- 2;
\item[{\em 3)}] in the unlikely case of high mass loss the \Heiiuv\ intensity is
maintained for up to $\sim$ 4 Myr;
\item[{\em 4)}] for models without mass loss, the equilibrium values of 
$I(1640)/I(\hb)$ are of the order of $\sim$ 0.1 -- 0.6 for the IMF
range considered here.
For comparison, $I(1640)/I(\hb) \sim$ 1.0 -- 3.2 for a power
law spectrum $L_\nu \propto \nu^{- \alpha}$ and typical values 
$\alpha \sim$  1 -- 1.8 for quasars (Elvis \etal\ 1994, Zheng \etal\ 1997), 
and the largest intensities
in (rare) metal-poor \hii\ regions are $I(1640)/I(\hb) \la$ 0.15, as
expected from their measured \Heiiopt\ intensity.
\end{itemize}

The equivalent widths for the strongest H and \heii\ lines and their
dependence on the IMF and stellar tracks are shown in Figs.\ \ref{fig_w1}
and \ref{fig_w2}.
Maximum equivalent widths of $\sim$ 700 and 3200 \AA\ are predicted for
the optical lines of \hb\ and \ha\ (not shown) respectively.
W(\lya) reaches values up to $\sim$ 2100 \AA\ at the ZAMS, but is 
potentially affected or destroyed by radiation transfer or other 
effects (cf.\ Tenorio-Tagle \etal\ 1999).
The maximum equivalent width of the \heii\ $\lambda\lambda$ 1640 and 4686 
recombination lines on the ZAMS is predicted to be $\sim$ 100 and 150 \AA\ 
respectively.

What variations are typically obtained for different IMF slopes ?
Varying $\alpha$ between 1. and 3. one finds the following changes
for bursts at age 0 Myr with respect to the Salpeter slope:
for models A and B $I(1640)/I(\hb)$ (and the other \heii\ intensities) varies between 
   $\sim$ -- 60 \% and + 50 \%,
$W({\rm Ly-}\alpha)$ changes by not more than $\sim$ 20 \%, and
$W($\Heiiuv$)$  varies between $\sim$ -- 50 \% and + 60 \%.
For obvious reasons, populations most biased towards massive stars
(i.e.\ large values of \mlow) are the least sensitive to the
exact IMF slope (cf.\ Bromm \etal\ 2001). E.g., the above quantities
vary by $\la$ 20 \% for model C.

Much larger \lya\ and \Heiiuv\ equivalent widths ($\sim$ 3100 and 1100 \AA\
respectively) have been predicted for a ZAMS population of exclusively massive 
stars by Bromm \etal\ (2001).
This overestimate is due to two effects acting in the same direction:
their simplified use of the spectrum of a 1000 \msun\ star with the hardest 
spectrum as representative for the entire population\footnote{For a Salpeter 
IMF from 300 -- 1000 \msun\ this leads to an 
overestimate of $L(1640)$ by a factor of 2 (cf.\ Sect.\ \ref{s_single}).}, 
and most importantly the neglect of nebular continuous emission, which 
--- for the case of a Salpeter IMF from 300 to 1000 \msun\ ---
contributes $\sim$ 75 \% of the total light at $\lambda$ 1640 and thus 
strongly reduces the \Heiiuv\ equivalent width (cf.\ above).
Accounting for these effects yields W(\lya) $\sim$ 2000 and W(1640) $\sim$
120 \AA\ for their IMF.

Our calculation of the nebular continuous emission assumes density bounded
objects, i.e.\ 
negligible escape of Lyman continuum photons out of the ``Pop III galaxies''
($f_{\rm esc} \ll 1$).
If this is not the case, one likely has a situation where nevertheless all
\heii\ ionising photons are absorbed --- i.e.\ $L($\Heiiuv$)$ remains identical ---
while some H ionising photons escape thereby reducing the continuum
emission (proportional to \Qh\ (Eq.\ \ref{eq_lcont}).
Although the escape fraction of ionising photons in distant galaxies
or the first building blocks thereof remains badly known,
there are theoretical and observational indications for rather low
escape fractions, of the order of $\sim$ 10 \% 
(see e.g.\ Leitherer \etal\ 1995, Dove \etal\ 2000, Steidel \etal\ 2001,
Hui \etal 2001). For such moderate an escape fraction, nebular continuous
emission remains thus dominant when strong emission lines are expected,
leading to \heii\ equivalent widths not much larger than predicted
here. 
Estimates for small Pop III halos suggest, however, large escape fractions
(Oh et al.\ 2001b). 
In short, the caveat regarding the uncertainty on $f_{\rm esc}$, which affects both 
emission line luminosities and nebular continuum emission, must be kept in mind.
A more detailed treatment including geometrical effects, radiation
transfer etc.\ (cf.\ Ciardi \etal\ 2001) are beyond the scope of our estimates.

Regarding the equivalent widths, we conclude that 
in any case the maximum values of $W($\lya$)$ and $W($\Heiiuv$)$ of
metal-free populations are $\sim$ 3--4 and $\sim$ 80 times larger
than the values predicted for very metal-poor stellar populations
($Z=1/50$ \zsun) with a ``normal'' IMF including stars up $\sim$ 100
\msun. When measured in the rest-frame, these will further be amplified by 
a factor $(1+z)$, where $z$ is the redshift of the source.

\subsection{Star formation indicators}

In the case of constant star formation, the ionising properties tend
rapidly (typically over $\sim$ 5 -- 10 Myr depending on the IMF) to their 
equilibrium value, which is then proportional to the star formation rate
(SFR).
In particular, recombination line luminosities $L_l$ are then simply 
given by
\begin{equation}
  L_l = c_l \, (1-f_{\rm esc}) \, Q_i 
                        \left(\frac{\rm SFR}{\msun {\rm yr}^{-1}}\right)
            = f_l \left(\frac{\rm SFR}{\msun {\rm yr}^{-1}}\right).
\label{eq_sfr}
\end{equation}
The predicted values $f_l$ (in erg s$^{-1}$) for \Heiiuv\ and \ha\ 
for the IMFs considered here are given in Table \ref{tab_sfr}
for the case of a negligible escape fraction of ionising photons
($f_{\rm esc} = 0$).
For other lines, 
or to compute the $Q_i$ not listed explicitly,
the corresponding value can easily be derived using 
Eq.\ \ref{eq_lines} and the emission coefficients in Table 
\ref{tab_lines}. Also listed are the \hei\ ionising production
and the photon flux in the Lyman Werner band of H$_2$ for 
a star formation rate of 1 \msun\ yr$^{-1}$.
For comparison with SFR indicators used for ``normal'' galaxies
we also indicate the mass conversion factor $c_M$ expressing
the relative masses between our model IMF and a ``standard'' Salpeter
IMF with \mlow\ $=$ 0.1 and \mup\ $=$ 100 \msun\ used
frequently throughout the literature, i.e.\ 
$c_M = \int_{0.1}^{100} M \Phi(M) dm / \int_{\mlow}^{\mup} M \Phi(M) dm$.

\begin{table*}
\caption{Line luminosities, supernova rates, and photon and metal production 
for constant star formation models normalised to SFR $=$ 1 \msun\ yr$^{-1}$}
\label{tab_sfr}
\begin{tabular}[htb]{lllllllllllllllll}
ID & $c_M$ & $f_{\ha}$ & $f_{1640}$ & \Qhe & \Qhtwo & $\bar{E_Q}$ & SNR & $M_{\rm ej}$ 
       & $M_{\rm ej}({\rm C})$&  $M_{\rm ej}({\rm O})$&  $M_{\rm ej}({\rm Si})$& $\eta$ \\ 
         & & \multicolumn{2}{c}{[erg s$^{-1}$]} & 
             \multicolumn{2}{c}{[(ph s$^{-1}$)/(\msun\ yr$^{-1}$)]} 
         & [eV] & [\msun$^{-1}$] & \multicolumn{4}{c}{[\msun/(\msun\ yr$^{-1}$)]} \\
\smallskip 
\\ \hline
A &  2.55 & 7.88e+41 & 1.91e+40 & 3.15e+53 & 3.17e+53 & 26.70 & 0.019 & 0.0076 & 0.0013 & 0.0030& 0.0005 & 0.065 \\
B &  2.30 & 1.03e+42 & 9.98e+40 & 4.34e+53 & 3.29e+53 & 27.86 & 0.016 & 0.030  & 0.0023 & 0.014 & 0.0045 & 0.022 \\
C &  14.5 & 3.32e+42 & 8.38e+41 & 1.54e+54 & 4.52e+53 & 29.69 & 0.0016& 0.14   & 0.0065 & 0.067 & 0.025  & 0.016 \\
\smallskip \\                                                         
D &  2.30 & 1.14e+42 & 3.12e+41 & 5.26e+53 & 3.36e+53 & 30.55 & 0.016  \\
E &  12.35 &4.10e+42 & 2.33e+42 & 2.21e+54 & 3.65e+53 & 35.15 & 0.0013 \\
\smallskip \\
 \multicolumn{6}{l}{Solar metallicity tracks:}   \\                   
ZS & 2.55 & 3.16e+41 & 5.12e+39 & 5.56e+52 & 3.60e+53 & 20.84 & 0.019 & 0.038$^\star$ &        &       &        & 0.0036$^\star$\\
 \multicolumn{6}{l}{$Z=1/50$ \zsun\ tracks:}   \\               
ZL & 2.47 & 6.92e+41 & 1.70e+39 & 1.55e+53 & 5.31e+53 & 21.95 & 0.019 & 0.023  & 0.0025 & 0.014 & 0.0014 & 0.014 \\
 \\ \hline
\multicolumn{13}{l}{$^\star$ Including stellar wind mass loss, SN Ibc, and SNII} 
\\ 
\end{tabular}
\end{table*}


Compared to the ``standard'' \ha\ SFR indicator for solar metallicity
objects (cf.\ Kennicutt 1998, Schaerer 1999) $L(\ha)$ is $\sim$ 2.4 
times larger for a Pop III object with the same IMF, given the
increased \Qh\ production (cf.\ TS00 and above).
Obviously, for IMFs more weighted towards high mass stars, 
the SFR derived from \ha\ (or other H recombination lines)
is even more reduced compared to using the standard Pop I SFR indicator.
If detected, \Heiiuv\ can also be used as a star formation indicator
as already pointed out by TGS01.
Their corresponding value is $f_{1640}=8.4 \, 10^{40} \times f_{\rm evol}$,
with $f_{\rm evol} \sim 0.4$ (2.) accounting in an approximate way for
stellar evolution effects without (with) mass loss.
Including the full sets of evolutionary tracks we find $f_{1640} = 1.91 \, 10^{40}$,
for the 1--100 \msun\ IMF, i.e.\ less \heii\ emission than TGS01.
Varying the IMF slope $\alpha$ between 1. and 3. leads to changes
of $f_{\ha}$ ($f_{1640}$) by $\sim \pm$ 0.5 (0.9) dex for models A and B,
and to minor changes ($\la$ 0.1 dex) for model C.
With the fits given in Table \ref{tab_fits} the line luminosities
can easily be computed for arbitrary IMFs.

In passing we note that the He$^+$ ionising flux per unit SFR used by 
Oh \etal\ (2001a) to estimate the expected number of sources with 
detectable \heii\ recombination lines has been overestimated:
for a constant SFR and a Salpeter IMF up to 100 \msun\ the hardness 
\Qhep/\Qh\ $\sim$ 0.005 is a factor 10 lower than their 
$Q$ value adopted\footnote{Oh \etal\ (2001a) adopt the ZAMS 
population of TS00 as representative of a stellar population
at equilibrium with a given SFR.}.
However, the Lyman continuum production is increased over the
value they adopted, leading thus to a net reduction of 
their flux $J_{\rm 1640}^{\rm stars}$ by a factor 1.9
(4.8) for a Salpeter IMF with \mlow\ = 1 (0.1) and \mup=100 \msun\
(cf.\ Tables \ref{tab_sfr} and \ref{tab_lines}).
Further issues regarding the photometric properties and the detectability 
of Pop III sources will be discussed in Pell\'o \& Schaerer (2001).

\subsection{Photon and heavy element production}

Various other properties related to the photon and heavy element production
are given in Table \ref{tab_sfr} for the case of constant star formation 
normalised to  SFR $=$ 1 \msun\ yr$^{-1}$.
Columns 5 and 6 give \Qhe, the He$^0$ ionising flux, and the
photon flux in the Lyman-Werner bands \Qhtwo\ respectively, both in units 
of (photon s$^{-1}$)/(\msun\ yr$^{-1}$).
The average energy $\bar{E_Q}$ (in eV) of the Lyman continuum photons
is in col.\ 7.

To estimate the metal production of Pop III objects we have used the
SN yields of Woosley \& Weaver (1995) for progenitor stars between
8 and 40 \msun\ and the results of Heger \etal\ (2000) and Heger \& Woosley (2001) 
for very massive objects of pair creation SN originating from stars 
in the mass range $\sim$ 130 to 260 \msun\ (cf.\ also Ober \etal\ 1993).
As the pair creation SN models of Heger and collaborators are calculated 
starting with pure helium cores, we have adopted the relation 
$M_\alpha \sim 0.5 \, M_{\rm ini}$ 
from Ober \etal\ (1983) to calculate the initial mass $M_{\rm ini}$.
We neglect the contribution from longer lived intermediate mass stars 
(1 $\la M/\msun \la$ 8).
Non-rotating stars more massive than $\sim$ 260 \msun\ are expected
to collapse directly to a black hole producing no metals (cf.\
Rackavy et al.\ 1967, Bond \etal\ 1984).

The SN rate SNR including ``normal'' SN and pair creation SN is given
in col.\ 8 of Table \ref{tab_sfr} (in units of SN per solar mass of stars
formed).
The total mass of metals $M_{\rm ej}$ produced including all elements heavier than 
He
is given in col.\ 9.
Columns 10, 11, and 12 indicate the contribution of C, O and Si.
The ejected masses are given in units solar mass per unit SFR.
Note that the models C and E with a lower mass cut-off \mlow\ $=$ 50 \msun\ 
of the IMF represent cases with solely pair creation SN (no SN type II),
models A (plus the cases ZS and ZL with non-zero metallicity) include
only SNII, and models B and D include both SN types.
Finally the dimensionless ``conversion efficiency'' of ionising photons to rest mass
(cf.\ Madau \& Shull 1996)
\begin{equation}
  \eta \equiv \frac{\int_0^\infty \int_{\nu_H}^{\infty} L_\nu d\nu dt}{M_{\rm ej} \, c^2}
       =      \frac{\Qh * \bar{E_Q}}{M_{\rm ej} \, c^2},
\end{equation}
with the spectral luminosity $L_\nu$, is given in col.\ 13.
It is useful to remind that, since both the metal and photon production is
dominated by massive stars, $\eta$ is independent of the IMF at masses $M \la$ 5--8 \msun,
and depends little on the exact shape of the IMF.

The SNR per unit mass depends little on the upper mass cut-off, as most 
of the mass resides in low mass stars. For cases C and E with 
\mlow\ = 50 \msun\ the SNR is reduced, since pair creation SN
from stars of a limited mass range ($\sim$ 130 -- 260 \msun) are the
only explosive events.

As evident from the yields of Woosley \& Weaver (1995) and shown in 
Table \ref{tab_sfr}, Population III stars convert --
for ``standard'' IMFs -- less of
their initial mass into metals than stars of non-zero 
metallicity\footnote{$M_{\rm ej}$ and $\eta$ given in Table 
\ref{tab_sfr} for solar metallicity (model ZS) have been calculated with
the code of Cervi\~no \etal\ (2000) taking into account stellar 
wind mass loss, SN Ibc, and SNII.
Neglecting mass loss and assuming only type II SN we obtain 
$M_{\rm ej} = $ 0.027 \msun\ and $\eta$ = 0.005, differing
by $\sim$ 40 \% from the tabulated value.}.
However, for IMFs favouring more strongly massive stars, allowing in particular
for pair creation SN ejecting up to half of their initial mass,
the metal production per unit stellar mass can be similar to or larger
than for solar metallicity.

The photon conversion efficiency of rest mass to ionising continuum,
$\eta$, of metal-poor and metal-free populations is increased
by factors of $\sim$ 3 - 18 compared to solar metallicity.
In great part this is due to increased ionising photon production.
The underlying changes of all quantities on which $\eta$ depends 
are listed in Table \ref{tab_sfr}.
The increase of $\eta$ from \zsun\ to 1/50 \zsun \footnote{
Assuming \mup\ $=$ 100 \msun\ as for models A and ZS reduces
\Qh\ by 16 \%.} indicates already
a substantially larger photon to heavy element production which must
have occurred in the early Universe.

In passing we note that our value of $\eta \sim 0.36 \%$ (or 0.5 \% 
neglecting stellar mass loss and SNIbc) is larger than the value of 
0.2 \% given by Madau \& Shull (1996), which, given the quoted
assumptions appears to be erroneous. 
This can be easily seen by noting the good agreement between
our photon production calculation and other calculations 
in the literature (e.g.\ the H$\alpha$ SFR calibration of
Kennicutt 1998, Glazebrook \etal\ 1999).
In addition one can easily estimate the metal production
from noting that $M_{\rm ej}(M) = f M_{\rm ini}$, with 
typically $ f \sim 0.1$ (Woosley \& Weaver 1995) and integrating 
over the IMF, which confirms our calculations.

Finally the predicted ejecta of carbon, oxygen, and silicon show
some interesting variations with the IMF.
From the ``standard'' IMF (A) to an IMF producing exclusively very 
massive stars (model C), 
the oxygen/carbon ratio (O/C) $\sim$ 0.8 (O/C)$_\odot$
increases by a factor $\sim$ 4,
while increases from Si/C $\sim$ 1.8 (Si/C)$_\odot$ by a factor 
of $\sim$ 10!
It is interesting to compare these values with current measurements
of abundance ratios in the Lyman-$\alpha$ forest, which 
indicate an overabundance of Si/C $\sim$ 2 -- 3 (Si/C)$_\odot$ 
(Songaila \& Cowie, 1996, Giroux \& Shull 1997), but possibly 
up to $\sim$ 10 times solar, depending on the adopted ionising 
UV background radiation field (Savaglio \etal\ 1997, Giroux \& Shull 1997).
If the ejecta from Pop III objects are responsible for the bulk of metals
in the Lyman-$\alpha$ forest, it would appear that IMFs strongly 
favouring massive stars seem to be excluded by studies
finding a modest Si/C overabundance.
The use of more detailed chemical evolution models including also
intermediate mass stars (cf.\ Abia \etal\ 2001) and 
additional studies on abundances in the Ly$\alpha$ forest
should hopefully provide stronger constraints on early nucleosynthesis
and the IMF of Pop III stars.
 
\section{Summary and discussion}
\label{s_summary}

The main aim of present study was to construct realistic models
for massive Population III stars and stellar populations to
study their spectral properties, including their dependence
on age, star formation history, and most importantly on the
IMF, which remains very uncertain for such objects.

We have calculated extensive sets of non-LTE model atmospheres
appropriate for metal-free stars with the {\em TLUSTY} code (Hubeny 
\& Lanz 1995) for plane parallel atmospheres.
The comparison with non-LTE models including stellar winds,
constructed with the comoving frame code {\em CMFGEN} of Hillier
\& Miller (1998), shows that even in the presence of some
putative (weak) mass loss, the ionising spectra of Pop III
stars differ negligibly from those of plane parallel models,
in stark contrast with Pop I stars (Gabler \etal\ 1989, Schaerer
\& de Koter 1997).
As already discussed by Tumlinson \& Shull (2000), the main salient
property of Pop III stars is their increased ionising flux, especially
in the He$^+$ continuum ($>$ 54 eV).

The model atmospheres have been introduced together with recent 
metal free stellar evolution tracks from the Geneva and Padova groups (Feij\'oo 
1999, Desjacques 2000, Marigo \etal\ 2000) and older Pop III tracks
assuming strong mass loss (Klapp 1983, El Eid \etal\ 1983) in the evolutionary 
synthesis code of Schaerer \& Vacca (1998) to study the temporal
evolution of individual Pop III stars and stellar populations.

The main results obtained for {\em individual Pop III stars} are the following
(Sect.\ \ref{s_single}):
\begin{itemize}
\item Due to their redward evolution off the zero age main sequence (ZAMS)
  the spectral hardness measured by the He$^+$/H ionising flux is decreased 
  by a factor $\sim$ 2 when averaged over their lifetime.
  If such stars would suffer strong mass loss, their spectral appearance
  could, however, remain similar to that of their ZAMS position.
\item The ionising photon fluxes of H, He$^0$, He$^+$, and the 
  flux in the Lyman-Werner band dissociating H$_2$ have been tabulated
  and detailed fit formulae are given for stars from 5 to 1000 \msun.
  These can in particular be used to calculate with arbitrary IMFs 
  the properties of ZAMS populations or cases of constant SFR.
  For comparison the same quantities are also given for solar metallicity
  and 1/50 \zsun.
\end{itemize}

The integrated spectral properties including line emission from H, \hei, \heii\ 
recombination lines (treated e.g.\ by TS00, Bromm \etal\ 2001) and 
nebular continuous emission --- neglected in all earlier studies --- 
have been calculated for instantaneous bursts and constant formation rates. 
Various assumptions on the mass cut-offs of the IMF including a ``standard'' 
IMF from 1 to 100 \msun\ or IMFs forming exclusively very massive stars
have been considered (see Table \ref{tab_models}).

The main results regarding {\em integrated stellar populations} are as follows
(Sect.\ \ref{s_pops}):
\begin{itemize}
\item For young bursts and the case of a constant SFR, nebular continuous
  emission dominates the spectrum or is at least important to properly predict
  it redward of Lyman-$\alpha$ if the escape fraction $f_{\rm esc}$ of ionising
    photons out of the observed region is small or negligible.
  In consequence predicted emission line equivalent widths are considerably
  smaller than found in earlier studies, whereas the detection of the continuum
  is eased. 
  Maximum equivalent widths of W(Ly$\alpha$) $\sim$ 1700 and W(\Heiiuv) $\sim$ 
  120 \AA\ are predicted for a Salpeter IMF with \mup\ $\sim$ 500 -- 1000 \msun.
  Nebular line and continuous emission strongly affect the broad band photometric
  properties of Pop III objects and (see Schaerer \& Pell\'o 2001, Pell\'o \& Schaerer
  2001).
  Detailed treatment models including geometrical effects, radiation
   transfer etc.\ (cf.\ e.g.\ Ciardi \etal\ 2001) will be useful to constrain
   $f_{\rm esc}$ more precisely.
\item Due to the redward stellar evolution and short lifetimes of the most massive
  stars, the hardness of the ionising spectrum decreased rapidly, leading to the
  disappearance of the characteristic \heii\ recombination lines after $\sim$ 3
  Myr in instantaneous bursts.
\item \Heiiuv, \ha\ (and other) line luminosities usable as indicators of 
  the star formation rate are given for the case of a constant SFR
  (Table \ref{tab_sfr})).
  For obvious reasons such indicators depend strongly on the IMF.
\item Accounting for SNII and pair creation supernovae we have also calculated
  supernova rates, the metal production, and the conversion efficiency
  of ionising photons to rest mass, $\eta$ (Table \ref{tab_sfr}).
  For a ``standard'' IMF with \mup\ $=$ 100 \msun\ the increased photon production 
  and the reduced metal yields of Pop III stars lead to an increase of $\eta$ by 
  a factor of 5 to 18 when compared to low (1/50 \zsun) or solar metallicity.
  In the presence of very massive stars ($M \sim$ 130 -- 260 \msun) leading to
  pair creation SN, somewhat lower values of $\eta \sim$ 1.6 to 2.2 \% are obtained.
  In any case metal-poor and metal-free populations are much more efficient
  producers of ionising photons 
  ``per baryon'' (per unit heavy element rest mass energy)
  than at solar metallicity, where
  $\eta \sim$ 0.36 \%.
\item Finally the ejected masses of C, O, and Si of Pop III objects have been calculated.
  The O/C abundance and even more so Si/C increase to highly supersolar values if
  very massive stars leading to pair creation SN form.
  How far measurements of such abundance ratios, e.g.\ in the Lyman-$\alpha$ forest,
  can be used to constrain the IMF of the first stellar objects, remains currently
  unclear, as also discussed by Abia \etal\ (2001).
\end{itemize}

The present model set should also be useful to explore the feasibility
of future direct observations of Pop III ``galaxies'' and to optimise
search strategies for such objects. We have started undertaking such 
studies (Schaerer \& Pell\'o 2001, Pell\'o \& Schaerer 2001).

Tumlinson \etal\ (2001) speculate that some emission line objects 
found in deep Lyman-$\alpha$ surveys could actually be \Heiiuv\
emitters such as the metal-free galaxies discussed here.
Recently Oh \etal\ (2001a) have estimated the number of Pop III sources and
mini quasars emitting strong \heii\ lines. Although still speculative, 
their calculations predict sufficient sources for successful detections
with deep, large field observations foreseeable with the Next Generation
Space Telescope or future instruments on ground-based 10m class telescopes.
Pilot studies now also start to address the question of dust formation and
obscuration in the early Universe (Todini \& Ferrara 2001), which remain 
a concern for observations.
In any case the upcoming decade should bring a great wealth of new information
on the early Universe, and possibly already the first {\em in situ} detection 
of the long sought Population III objects.

\begin{acknowledgements}
This project is partly supported by INTAS grant 97-0033.
I would like to thank Andr\'e Maeder, Georges Meynet,
Vincent Desjacques, and Paola Marigo for sharing new and 
partly unpublished stellar evolution tracks.
Best thanks to Ivan Hubeny and Thierry Lanz for making {\em TLUSTY}
public, to John Hillier for sharing his sophisticated
{\em CMFGEN} atmosphere code, and to Miguel Cervi\~no 
for test calculations with his evolutionary synthesis code.
I warmly thank Roser Pell\'o for stimulating discussions,
as well as Tom Abel, Andrea Ferrara, Alexander Heger, Fumitaka Nakamura 
and Patrick Petitjean for useful comments on various issues.
\end{acknowledgements}


\end{document}